\documentclass[11pt,letterpaper]{article}

\usepackage{jheppub}
\usepackage{xcolor}
\usepackage{graphicx}
\usepackage{xspace}
\usepackage{wrapfig,enumerate,slashed}

\usepackage[bottom]{footmisc}
\usepackage{amsmath}
\usepackage{wasysym} 
\usepackage{graphicx}
\usepackage{color}
\usepackage{orcidlink}
\usepackage{hyperref}
\usepackage{subfigure}

\usepackage{soul}

\renewcommand{\d}{{\mathrm{d}}}

\def\mw{m_{\rm W}}
\def\mz{m_{\rm Z}}
\def\mh{m_{\rm H}}

\def\vev{{\it v}}

\def\Zf{Z}
\def\div{\Delta_\epsilon}
\def\deltaCT{{\delta_\epsilon}}

\newcommand{\nn}{\nonumber}
\newcommand{\be}{\begin{equation}}
\newcommand{\ee}{\end{equation}}
\newcommand{\bear}{\begin{eqnarray}}
\newcommand{\eear}{\end{eqnarray}}
\newcommand{\mL}{\mathcal{L}}
\newcommand{\mO}{\mathcal{O}}

\def\1loop{one-loop}

\def\greenfR{\hat{\Gamma}}

\def\greenfL{\Gamma^{\rm Loop}}
\def\greenfC{\Gamma^{\rm CT}}
\newcommand{\MSb}{$\overline{\text{MS}}${}}
\newcommand{\mg}{\texttt{MadGraph5\_aMC@NLO}}

\begin{document}
\allowdisplaybreaks

%%%%%%%%%%%%%%%%%%%%%%%%%%%%%%%%%%%%%%%%%%%%%%%%%%%%%
\title{Bosonic multi-Higgs correlations beyond leading order}
%%%%%%%%%%%%%%%%%%%%%%%%%%%%%%%%%%%%%%%%%%%%%%%%%%%%%

%%%%%%%%%%%%%%%%%%%%%%%%%%%%%%%%%%%%%%%%%%%%%%%%%%%%%
\author[a]{Anisha\orcidlink{0000-0002-5294-3786},}
\author[b]{Daniel Domenech\orcidlink{0000-0001-5967-9044},}
\author[a]{Christoph Englert\orcidlink{0000-0003-2201-0667},}
\author[b]{Maria J. Herrero\orcidlink{0000-0002-2322-1629},}
\author[c]{and Roberto A. Morales\orcidlink{0000-0002-9928-428X}}
%%%%%%%%%%%%%%%%%%%%%%%%%%%%%%%%%%%%%%%%%%%%%%%%%%%%%

%%%%%%%%%%%%%%%%%%%%%%%%%%%%%%%%%%%%%%%%%%%%%%%%%%%%%
\affiliation[a]{School of Physics and Astronomy, University of Glasgow, Glasgow G12 8QQ, United Kingdom}
\affiliation[b]{Departamento de F\'isica Te\'orica and Instituto de F\'isica T\'orica, IFT-UAM/CSIC, Universidad Aut\'onoma de Madrid, Cantoblanco, 28049 Madrid, Spain}
\affiliation[c]{IFLP, CONICET - Dpto. de F\'isica, Universidad Nacional de La Plata, C.C. 67, 1900 La Plata, Argentina}
%%%%%%%%%%%%%%%%%%%%%%%%%%%%%%%%%%%%%%%%%%%%%%%%%%%%%
\emailAdd{anisha@glasgow.ac.uk}
\emailAdd{daniel.domenech@uam.es}
\emailAdd{christoph.englert@glasgow.ac.uk}
\emailAdd{maria.herrero@uam.es}
\emailAdd{roberto.morales@fisica.unlp.edu.ar}
%%%%%%%%%%%%%%%%%%%%%%%%%%%%%%%%%%%%%%%%%%%%%%%%%%%%%

%%%%%%%%%%%%%%%%%%%%%%%%%%%%%%%%%%%%%%%%%%%%%%%%%%%%%
\abstract{The production of multiple Higgs bosons at the LHC and beyond is a strong test of the mechanism of electroweak symmetry breaking. Taking inspiration from recent experimental efforts to move towards limits on triple Higgs production at the Large Hadron Collider, we consider generic bosonic deviations of $HH$ and $HHH$ production from the Standard Model in the guise of Higgs Effective Field Theory.  Including one-loop radiative corrections within the HEFT and going up to ${\mathcal{O}}(p^4)$ in the momentum expansion, we provide a detailed motivation of the parameter range that the LHC (and future hadron colliders) can explore, through accessing non-standard coupling modifications and momentum dependencies that probe Higgs boson non-linearities.  In particular, we find that radiative corrections can enhance the sensitivity to Higgs-self coupling modifiers and HEFT-specific momentum dependencies can vastly increase triple Higgs production thus providing further motivation to consider these processes during the LHC's high-luminosity phase. }
%%%%%%%%%%%%%%%%%%%%%%%%%%%%%%%%%%%%%%%%%%%%%%%%%%%%%

\preprint{}

\maketitle
%%%%%%%%%%%%%%%%%%%%%%%%%%%%%%%%%%%%%%%%%%%%%%%%%%%%%
\section{Introduction}
\label{sec:intro}
%%%%%%%%%%%%%%%%%%%%%%%%%%%%%%%%%%%%%%%%%%%%%%%%%%%%%
Although the Higgs discovery in 2012 has catapulted particle physics into an era of exploration of the TeV scale with direct phenomenological measurements, the consistency of LHC data with theoretical predictions of the Standard Model of Particle Physics (SM) is startling. Many aspects of nature are understood to be unsatisfactorily described by the SM, and it is therefore surprising that the SM's success lives on some ten years after the discovery of the Higgs boson. That said, the plethora of physics observations that informed the Higgs boson discovery can be attributed to the LEP electroweak precision programme and the theoretical requirement of spontaneous symmetry breaking. It is the latter that is still poorly understood, and while the question of how the flavour sector intersects with electroweak symmetry breaking is an important one, the very existence of the TeV scale is a pressing question in its own right. 

Clearly, we are missing a crucial point and given that the SM is by construction ultraviolet complete, clues will likely be provided by the experiments. It will fall onto theory to contextualize any such new physics signature with the questions that are left unanswered by the SM. Along these lines, the electroweak symmetry-breaking potential remains a motivated line of inquiry: The LHC multi-purpose experiments are becoming increasingly sensitive to tell-tale modifications in multi-Higgs boson rates that fingerprint the electroweak potential, but data are at such an early stage that large deviations would still go unnoticed. In fact, such deviations can be expected in many scenarios of physics beyond the SM (BSM) with relevance for physics at the TeV scale but also for the early universe (see e.g.~\cite{Grojean:2004xa,Bodeker:2004ws}), thus making the Higgs potential a high-value phenomenology target for the high-luminosity phase of the LHC and~beyond.

In this work, we perform a detailed precision investigation of the most general bosonic deformation of the TeV scale. By employing Higgs Effective Field Theory (HEFT), we obtain expectations of multi-Higgs $pp\to HH$ and $pp \to HHH$ production rates, beyond the leading order. This means that we will include one-loop radiative corrections by means of the relevant 1-Particle-Irreducible (1PI) Higgs functions and will consider up to chiral dimension 4 operators in the HEFT Lagrangian (for the relevant set of these operators, see for instance~\cite{Buchalla:2013eza, Brivio:2013pma}). By construction, we, therefore, obtain all relevant bosonic correlation modifications under the assumption of new physics around the TeV scale, in particular generalising the expectations of Standard Model Effective Field Theory (SMEFT). In particular, the triple Higgs final states are extremely rare in the SM at the LHC but have been highlighted as a target of a future hadron machine or BSM extensions~\cite{Chen:2015gva,Fuks:2015hna,Papaefstathiou:2015paa,Fuks:2017zkg,Kilian:2017nio,Abdughani:2020xfo,Papaefstathiou:2019ofh,Papaefstathiou:2020lyp,Papaefstathiou:2023uum,Stylianou:2023xit,Karkout:2024ojx}. Our analysis provides a theoretically robust interpretation framework to interpret large enhancements of the current and future LHC multi-Higgs programme, with direct relevance to astrophysically relevant observations~\cite{Alonso:2023jsi}.

This paper is organized as follows. In Sec.~\ref{sec:calc}, we briefly introduce Higgs Effective Field Theory and provide details of our (higher-order) calculation and implementation. In Sec.~\ref{sec:pheno}, we survey $HHH$ and $HH$ production in HEFT beyond leading order. We summarize our findings and provide an outlook in Sec.~\ref{sec:conc}. In the final appendix, we have included all the details of the 1PI Higgs functions involved in the present computation.

%%%%%%%%%%%%%%%%%%%%%%%%%%%%%%%%%%%%%%%%%%%%%%%%%%%%%
\section{HEFTy Multi-Higgs production beyond leading order}
%%%%%%%%%%%%%%%%%%%%%%%%%%%%%%%%%%%%%%%%%%%%%%%%%%%%%
\subsection{Elements of the Calculation}
\label{sec:calc}
\begin{figure}[!t]
\centering
\includegraphics[height=0.23\textwidth]{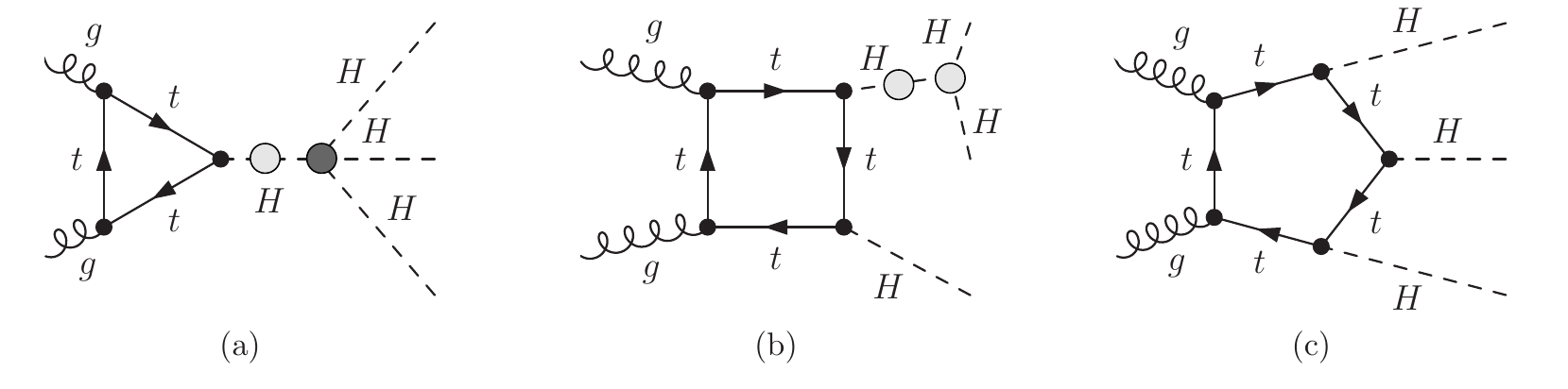}
\caption{Representative Feynman diagram topologies contribution to gluon fusion triple Higgs production: (a) triangle diagrams, (b) box topologies and (c) pentagon diagrams. The light-shaded regions refer to 1-PI {\emph{irreducible}} vertex functions, Eqs.~\eqref{eq:2pt},  \eqref{eq:3pt} and \eqref{eq:4pt} whereas the dark-shaded ones refer to the connected {\emph{reducible}} 4-point scalar vertex function, Eq. ~\eqref{eq:4pta}. The fermion-Higgs sector interactions and the fermion-gauge sector interactions are taken to be SM-like for the parameter choices of this work and also the employed renormalisation scheme.  Hence, these fermionic diagrams are identical to the SM ones.\label{fig:Feyn}}
\end{figure}

\begin{figure}[!b]
\centering
\includegraphics[height=0.23\textwidth]{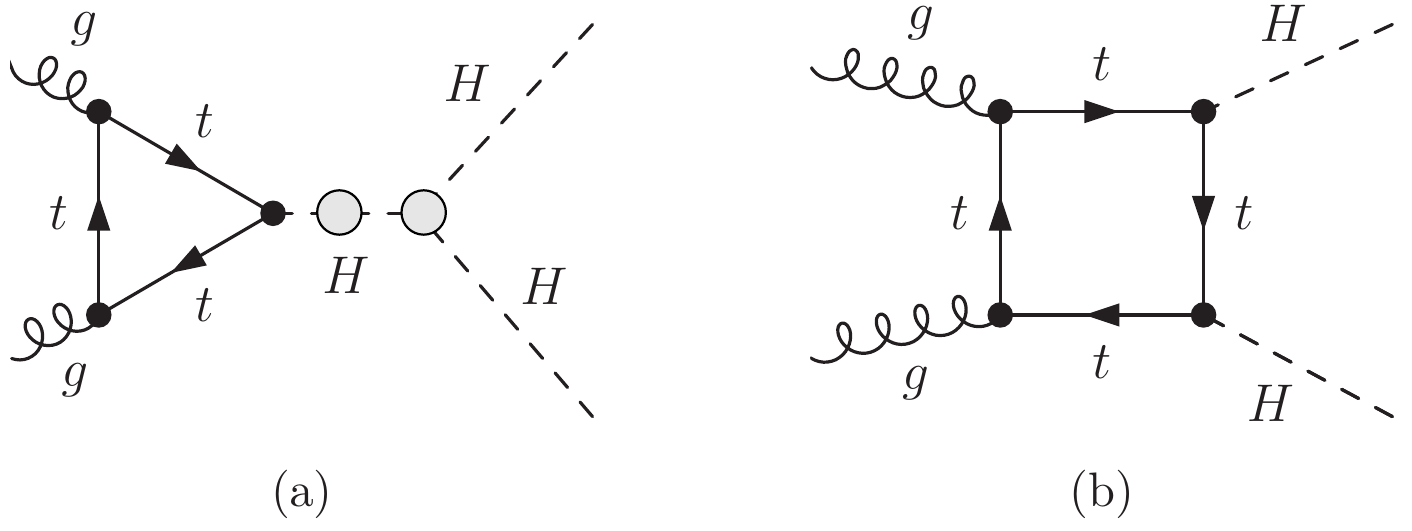}
\caption{Representative Feynman diagram topologies contribution to gluon fusion double Higgs production similar to Fig.~\ref{fig:Feyn}: (a) triangle diagrams, and (b) box topologies.  The light-shaded regions again refer to 1-PI {\emph{irreducible}} vertex functions. 
\label{fig:Feynhh}}
\end{figure}

%%%%%%%%%%%%%%%%%%%%%%%%%%%%%%%%%%%%%%%%%%%%%%%%%%%%%
In this work, we consider the two-loop factorisable contributions to $pp\to HH,HHH$, indicated in Figs.~\ref{fig:Feyn} and \ref{fig:Feynhh} for the two Higgs production and  the more complex case of triple Higgs production.  While the gluon fusion topologies for single Higgs production have a long history~\cite{Georgi:1977gs,Djouadi:1991tka,Graudenz:1992pv,Harlander:2002wh,Anastasiou:2015vya}, the two- and three-Higgs contributions have been considered in detail much later, in particular, the two-loop QCD contribution to Higgs pair production in the finite top mass limit have only been made available recently~\cite{Borowka:2016ypz,Baglio:2018lrj}. In this work, we will focus on the finite top limit, at leading order, whilst focusing on the Higgs interactions within HEFT beyond leading order considerations. This is a motivated approach as the NLO QCD interactions are known to be relatively insensitive to the electroweak details of the amplitude beyond the relevance of the top mass threshold~\cite{deFlorian:2019app}. Therefore, by reporting results in comparison with the SM expectation, we can be confident that most QCD-relevant aspects will generalize to the inclusion of the electroweak effects that we consider in this work.

Throughout, following Refs.~\cite{Herrero:2021iqt,Herrero:2022krh,Anisha:2022ctm}, we will use the on-shell renormalisation scheme for the physical states, the electric charge is defined from the full $ee\gamma$ on-shell three-point function at vanishing momentum transfer (the Thomson limit), and the HEFT coefficients are renormalized in the \MSb~scheme. The building blocks in for, e.g., $gg \to HHH$ are shown in Fig.~\ref{fig:Feyn} and for $gg \to HH$ in Fig. \ref{fig:Feynhh}.  More precisely, we consider only factorisable two-loop contributions in this work, which is a widely used approximation for Higgs sector extensions, see, e.g., the recent~\cite{Heinemeyer:2024hxa} and references therein.

%%%%%%%%%%%%%%%%%%%%%%%%%%%%%%%%%%%%%%%%%%%%%%%%%%%%%
%%%%%%%%%%%%%%%%%%%%%%%%%%%%%%%%%%%%%%%%%%%%%%%%%%%%%

%%%%%%%%%%%%%%%%%%%%%%%%%%%%%%%%%%%%%%%%%%%%%%%%%%%%%
%%%%%%%%%%%%%%%%%%%%%%%%%%%%%%%%%%%%%%%%%%%%%%%%%%%%%

The HEFT Lagrangian is organized by the chiral dimension into two parts, $\mL_2$ and $\mL_4$, corresponding to operators of $\mO(p^2)$ and $\mO(p^4)$, respectively.
The relevant operators for this work are (we are using here the notation of Ref.~\cite{Herrero:2022krh}):
\begin{subequations}
\label{eq:lag}
\bear
\mL_2 &=& \frac{1}{2} \partial_{\mu} H \partial^{\mu} H - \left(\frac12 \mh^2 H^2 + 
\frac12 \kappa_3 \frac{\mh^2}{v} H^3 + \frac{1}{8}\kappa_4 \frac{\mh^2}{v^2} H^4
 \right) \nn \\
&&+\frac{v^2}{4}  \left(1 + 2 a \frac{H}{v} + b \frac{H^2}{v^2}\right)
        \text{Tr}[D_{\mu} U^{\dagger} D^{\mu} U]
        - \frac{1}{2 g^2} \text{Tr}[\hat{W}_{\mu \nu} \hat{W}^{\mu \nu}] - \frac{1}{2 g'^2} \text{Tr}[\hat{B}_{\mu \nu} \hat{B}^{\mu \nu}]\,, \nn\\  %+ {\mL}_{GF} + {\mL}_{FP}  \nn\\
\eear
and
\bear        
        \mL_4 &=& \mO_{\square\square}+\mO_{H\square\square}+\mO_{HH\square\square} +\mO_{dd\square}+\mO_{Hdd\square} +\mO_{dddd}  \nn\\
 &&\hspace{3cm}+\mO_{Hdd}+\mO_{HHdd} +\mO_{ddW}+\mO_{HddW} +\mO_{ddZ}+\mO_{HddZ}  \,,
\eear
\end{subequations}
with $U=\exp(i\pi^a t^a/v)$ as the non-linear sigma field of $[SU(2)_L\times SU(2)_R]/SU(2)_{L+R}$ ($\pi^a$ are the would-be Nambu-Goldstone fields), thus preserving custodial isospin. Hypercharge is embedded $U(1)_Y\subset SU(2)_R$ so that the model describes a nonlinear realisation of electroweak symmetry breaking $SU(2)_L\times U(1)_Y\to U(1)_{\text{em}}$. The field strength tensors are defined as 
\begin{equation}
\begin{split}
\hat{W}_{\mu \nu} &= \partial_\mu \hat W_\nu  - \partial_\nu \hat W_\mu + i [ \hat W_\mu,  \hat W_\nu]\,,\\
\hat{B}_{\mu \nu} &= \partial_\mu \hat B_\nu  - \partial_\nu \hat B_\mu \,,~\text{with}\\
 \hat W_\mu  &= g W^a_\mu {\tau^a\over 2}\,,~
 \hat B_\mu  = g' B_\mu {\tau^3\over 2}
 \end{split}
 \end{equation}
 with weak and hypercharge couplings $g,g'$, respectively; $\tau^a$ are the Pauli matrices. Electroweak gauging in Eq.~\eqref{eq:lag} is achieved through the covariant derivative
\begin{equation}
D_{\mu}U=\partial_{\mu}U+i  \hat{W}_{\mu}U-i U \hat B_{\mu}\,.
\end{equation}
Notice that gauge-fixing and Faddeev-Popov terms were omitted for simplicity in $\mL_2$, in particular, we will implement the Feynman gauge in the numerical computations. Furthermore, we are assuming that fermionic interactions are the same as in the SM. Hence, we consider the new physics only in the bosonic sector. Concretely, in the chiral dimension two Lagrangian, it is parametrized by the $a$, $b$, $\kappa_3$ and $\kappa_4$ HEFT coefficients, and the chiral dimension four operators are collected in Tab.~\ref{tab:operators}. 

%
%
%%%%%%%%%%%%%%%%%%%%%%%%%%%%%%%%%%%%%%%%%%%%%%%%%%%%%
\begin{table}[!t]
	\centering
	\renewcommand{\arraystretch}{2.0}
	\small{	\begin{tabular}{|c|c||c|c|}
			\hline
			$\mathcal{O}_{\square \square}$ 
			& $a_{\square \square }  \frac{\square H \square H}{v^2}$ &
			$\mathcal{O}_{H\square \square}$ 
			& $a_{H\square \square}  \; \big(\frac{H}{v}\big) \;  \frac{\square H \square H}{v^2}$ \\
			\hline
			$\mathcal{O}_{Hdd}$ 
			& $a_{Hdd } \; \frac{\mh^2 }{v^2} \big(\frac{H}{v}\big) \; \partial^{\mu} H \partial_{\mu} H$ &
			$\mathcal{O}_{HHdd}$ 
			&$a_{HHdd}  \; \frac{\mh^2 }{v^2} \big(\frac{H^2}{v^2}\big) \; \partial^{\mu} H \partial_{\mu} H$ \\
			\hline
			$\mathcal{O}_{ddW}$ 
			& $a_{ddW } \; \frac{\mw^2 }{v^2} \big(\frac{H}{v}\big) \; \partial^{\mu} H \partial_{\mu} H$ &
			$\mathcal{O}_{HddW}$ 
			& $a_{HddW } \; \frac{\mw^2 }{v^2} \big(\frac{H^2}{v^2}\big) \; \partial^{\mu} H \partial_{\mu} H$ \\
			\hline
			$\mathcal{O}_{ddZ}$ 
			& $a_{ddZ } \; \frac{\mz^2 }{v^2} \big(\frac{H}{v}\big) \; \partial^{\mu} H \partial_{\mu} H$ &
			$\mathcal{O}_{HddZ}$ 
			& $a_{HddZ} \; \frac{\mz^2 }{v^2} \big(\frac{H^2}{v^2}\big) \; \partial^{\mu} H \partial_{\mu} H$ \\
			\hline
			$\mathcal{O}_{dd\square }$  &
			$a_{dd\square} \; \frac{1}{\vev^3} \,  \partial^\mu H\,\partial_\mu H\,\square H$ &
			$\mathcal{O}_{Hdd\square}$ 
			& $a_{Hdd\square } \; \frac{1}{v^3} \big(\frac{H}{v}\big)\; \partial^\mu H \, \partial_\mu H \, \square H $ \\
			\hline
			$\mathcal{O}_{HH\square \square}$ 
			& $a_{HH\square \square}  \; \big(\frac{H^2}{v^2}\big) \frac{\square H \square H}{v^2}$ &
			$\mathcal{O}_{dddd}$ 
			&$a_{dddd}  \; \frac{1}{v^4} \partial^\mu H\,\partial_\mu H\,\partial^\nu H\,\partial_\nu H$ \\
			\hline
			\end{tabular}}
	\caption{Relevant HEFT operators $\mathcal{O}_{i}$  with $a_{i}$ being the corresponding HEFT coefficients.} 
	\label{tab:operators}
\end{table}  
%%%%%%%%%%%%%%%%%%%%%%%%%%%%%%%%%%%%%%%%%%%%%%%%%%%%%
%

Once the on-shell renormalisation conditions are imposed, the counterterms and all the renormalized 1-PI vertex functions are written in terms of the renormalized parameters electric charge and physical masses $\mw=gv/2$, $\mz=\sqrt{g^2+g'^2}\,v/2$ and $\mh^2 =2 v^2 \lambda$.  From this Lagrangian, one can derive the interactions that are relevant for the calculation of the relevant (sub)amplitudes. The tree level contribution comes from both $\mL_2$ and $\mL_4$,  whereas the loop  contributions arise from the chiral dimension two Lagrangian only.  The counterterms come from the chiral dimension two Lagrangian, and also the $a_i$  coefficients of the chiral dimension four act as counterterms. The building blocks relevant for the $gg\to HHH$ ($gg\to HH$ follows from similar decomposition as shown in Fig.~\ref{fig:Feynhh}) are the irreducible two, three, and four-point vertex functions. The renormalized self-energy of the (iso-singlet) Higgs boson is given by
\begin{equation}
\begin{split}
\label{eq:2pt}
\parbox{2.7cm}{\includegraphics[width=2.7cm]{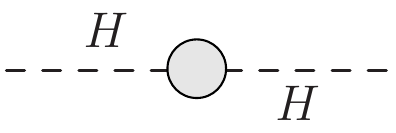}}
~&=-i\hat{\Sigma}_{HH}(q^2)\\ &= -i\Sigma_{HH}^{\rm Loop}(q^2) +i\left(\delta\Zf_H (q^2-\mh^2) -\delta\mh^2\right) +i\frac{2a_{\Box\Box}}{\vev^2}q^4\,,
\end{split} 
\end{equation}
such that 
\begin{equation}
\hat{\Sigma}_{HH}(\mh^2) = {\d \hat{\Sigma}_{HH} \over \d q^2}\bigg|_{q^2=\mh^2} = 0
\end{equation}
in the on-shell scheme. 

The electroweak vacuum expectation value (vev) is fixed through the gauge boson masses assuming custodial invariance as highlighted in Eq.~\eqref{eq:lag}. The renormalisation conditions are tabled in Refs.~\cite{Herrero:2021iqt,Herrero:2022krh}; it is worth highlighting that owing to the singlet nature of the Higgs boson in HEFT, any gauge dependence cancels explicitly. Although, at face value, HEFT is a much broader class of field theories (see in particular Ref.~\cite{Alonso:2016oah}), this fact together with similar cancellations of gauge dependencies in the gauge boson sector~\cite{Herrero:2021iqt,Herrero:2022krh}, lead to technical simplifications that are not present in, e.g., SMEFT. Furthermore, it is known that the HEFT approximates the resummation behaviour of SMEFT~\cite{Helset:2020yio}. Nonetheless, the latter can be obtained from the former through appropriate redefinitions, which in turn alludes to a less transparent power counting of HEFT and possibly large scheme dependencies (typically resolved in matching calculations).

The renormalized irreducible 3-point vertex function is parametrized as
\begin{align}
\label{eq:3pt}
\parbox{2.4cm}{\includegraphics[width=2.4cm]{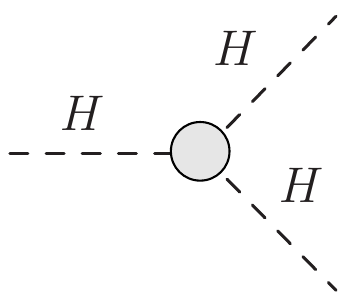}}
~&= i\greenfR_{HHH}(p_1,p_2,p_3) \nn  \\
 &= -3i\kappa_3\frac{\mh^2}{\vev} +i\greenfL_{HHH} -3i\kappa_3\frac{\mh^2}{\vev}\left(\frac{\delta\kappa_3}{\kappa_3}+\frac{\delta\mh^2}{\mh^2}-\frac{\delta\Zf_\pi}{2}-\frac{\delta\vev}{\vev}+\frac{3\delta\Zf_H}{2}\right) \nn \\
& +\frac{i}{\vev^3}\left(a_{dd\Box}(p_1^4+p_2^4+p_3^4)+2(a_{H\Box\Box}-a_{dd\Box})(p_1^2p_2^2+p_2^2p_3^2+p_3^2p_1^2) \right.\nn\\
&\left. \hspace{10mm}+(a_{Hdd}\,\mh^2+a_{ddW}\,\mw^2+a_{ddZ}\,\mz^2)(p_1^2+p_2^2+p_3^2)\right)\nn\,,\\
\end{align}
which is manifestly invariant under permutations of the incoming external momenta $p_i$. This shows a further motivation for the HEFT formalism that is rooted in its relation to the Lorentz structures that the different interactions induce. These are directly related to experimental measurements. SMEFT, in contrast, selects correlations in this space through {\emph{internal}} symmetry considerations. Similar to the 3-point vertex, the renormalized four-point function is given by
\begin{equation}
\begin{split}
\label{eq:4pt}
\parbox{4cm}{\includegraphics[width=4cm]{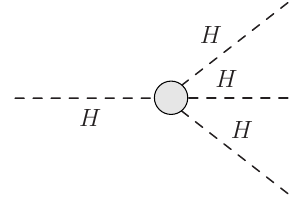}}
~&=i\greenfR_{HHHH}(p_1,p_2,p_3,p_4) \\ 
&\hspace{-2cm}= -3i\kappa_4\frac{\mh^2}{\vev^2} +i\greenfL_{HHHH} -3i\kappa_4\frac{\mh^2}{\vev^2}\left(\frac{\delta\kappa_4}{\kappa_4}+\frac{\delta\mh^2}{\mh^2}-2\left(\frac{\delta\Zf_\pi}{2}+\frac{\delta\vev}{\vev}\right)+2\delta\Zf_H\right) \\
&\hspace{-2cm}+ \frac{i}{\vev^4} \left(a_{Hdd\Box}(p_1^4+p_2^4+p_3^4+p_4^4-2(p_1^2p_2^2+p_1^2p_3^2+p_1^2p_4^2+p_2^2p_3^2+p_2^2p_4^2+p_3^2p_4^2)) \right. \\
&\hspace{-2cm}\left. +4a_{HH\Box\Box}(p_1^2p_2^2+p_1^2p_3^2+p_1^2p_4^2+p_2^2p_3^2+p_2^2p_4^2+p_3^2p_4^2) \right. \\
&\hspace{-2cm}\left. +
{2}(a_{HHdd}\,\mh^2+a_{HddW}\,\mw^2+a_{HddZ}\,\mz^2)(p_1^2+p_2^2+p_3^2+p_4^2) \right. \\
&\hspace{-2cm}      +4a_{dddd}\left((p_1+p_2)^2(p_1+p_3)^2+(p_1+p_2)^2(p_2+p_3)^2+(p_1+p_3)^2(p_2+p_3)^2 \right. \\
&\hspace{-0.2cm}\left.\left. -(p_1^2p_2^2+p_1^2p_3^2+p_1^2p_4^2+p_2^2p_3^2+p_2^2p_4^2+p_3^2p_4^2)  \right)\right)\,.
\end{split}
\end{equation}
While the irreducibly three-point interactions directly enter the box diagrams of the $gg\to HHH$ amplitude (as well as the triangle diagrams of $pp \to HH$), the triangle contributions to triple Higgs production feature the reducible (truncated) four-point contribution as indicated in Fig.~\ref{fig:Feyn} (a). These are represented as
\begin{multline}
\label{eq:4pta}
{i\overline{{G}}}_{HHHH}=
\parbox{3.5cm}{\includegraphics[width=3.5cm]{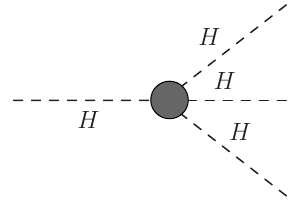}}
~=~\parbox{3.5cm}{\includegraphics[width=3.5cm]{h2.pdf}}
~+~\parbox{3.5cm}{\includegraphics[width=3.5cm]{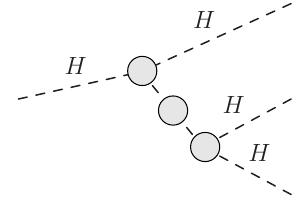}}\\
+~\text{permutations}\,.
\end{multline}
With these building blocks, one can expand the squared amplitude including the chiral dimension four effects as
\begin{equation}
\label{eq:ampexp}
|{\cal{M}} |^2= |{\cal{M}}_{d=2}|^2 + 2\, \text{Re}\left\{  {\cal{M}}_{d=2}{\cal{M}}_{d=4}^\ast \right\}\,,
\end{equation}
where the $d=2,4$ parts derive from the expansion of the representation of the normalized amplitude in Fig.~\ref{fig:Feyn} at the specified chiral dimension recorded in Eq.~\eqref{eq:lag} and Tab.~\ref{tab:operators}, respectively. In particular, triangle, box and pentagon contributions are obtained by multiplying with the relevant (off-shell) scalar currents, cf. Eqs.~\eqref{eq:2pt}-\eqref{eq:4pta}.

Following Refs.~\cite{Herrero:2021iqt,Herrero:2022krh,Anisha:2022ctm}, the relevant functions have been obtained through a combination of {\tt{FeynRules}}~\cite{Christensen:2008py,Alloul:2013bka}, {\tt{FeynArts}}, {\tt{FormCalc}}~\cite{Hahn:2000kx,Hahn:1998yk,Hahn:2000jm,Hahn:2006qw} interfaced with {\tt{Vbfnlo}}~\cite{Arnold:2008rz}. We have validated our implementation for SM parameter choices against \mg~\cite{Alwall:2014hca, Hirschi:2015iia}. Furthermore, analytical cross-checks have been performed to verify our implementation of the reducible 4-point function assuming the SM. Further details of the calculation of the renormalized 1PI Higgs functions are given in the appendix~\ref{1PIfunctions}.

%%%%%%%%%%%%%%%%%%%%%%%%%%%%%%%%%%%%%%%%%%%%%%%%%%%%%
\subsection{Phenomenology and Discussion}
\label{sec:pheno}
%%%%%%%%%%%%%%%%%%%%%%%%%%%%%%%%%%%%%%%%%%%%%%%%%%%%%
\newcommand{\ab}{~\text{ab}}
\newcommand{\fb}{~\text{fb}}
As with $gg\to HH$~\cite{Baur:2002rb,Dolan:2012rv}, the interplay of triangle, box, and pentagon topologies in $gg\to HHH$, driven by the finite top threshold,  is critical for the expected phenomenology. With a total cross section of around $50\ab$ in the SM at 14 TeV~\cite{deFlorian:2019app} (rising to $\sim 3.5\fb$ at a 100 TeV FCC-hh), with significant theoretical uncertainties, a differential analysis of the invariant $HH$ and $HHH$ momenta in triple Higgs production is severely restricted at the high luminosity (HL)-LHC. Therefore, we will limit ourselves to a discussion of inclusive rates that can be expected for $HH$ and $HHH$ production as a function of the HEFT coefficients. 

\begin{figure}[!t]
\centering
\includegraphics[width=0.49\textwidth]{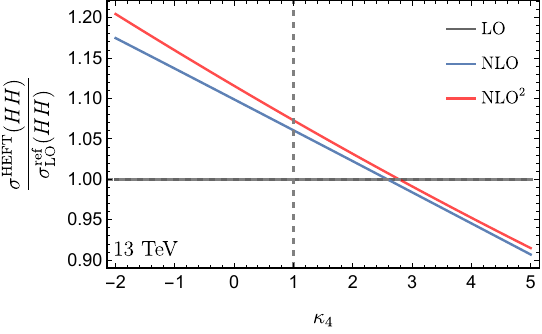}\hfill
\includegraphics[width=0.49\textwidth]{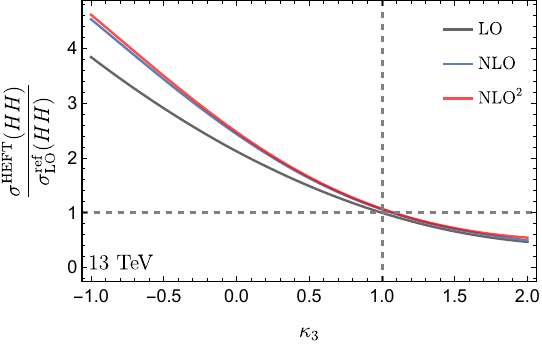}\\[0.2cm]
\includegraphics[width=0.49\textwidth]{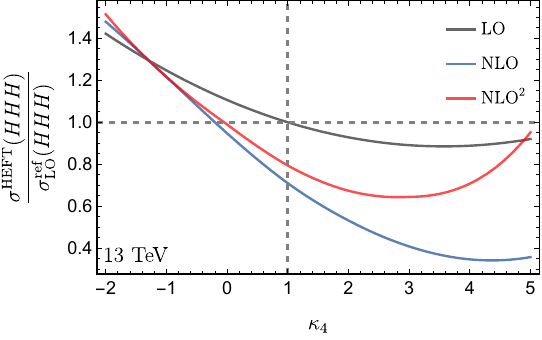}\hfill
\includegraphics[width=0.49\textwidth]{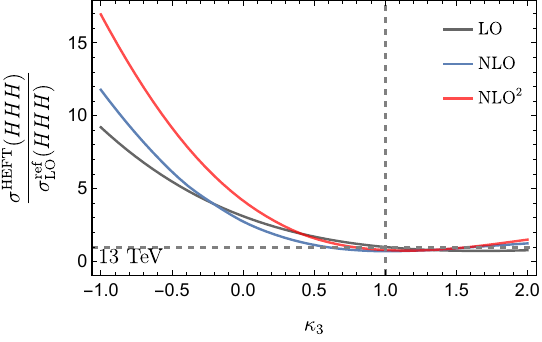}
\caption{\label{fig:kfac} Cross sections within HEFT relative to the reference LO expectation,  cf. Eq.~\eqref{eq:SMversusref1}.  Here we focus on sensitivities to $\kappa_3$ and $\kappa_4$ and set all NLO coefficients to $a_i=0$. To highlight regions of strong coupling, we also include the squared NLO contributions which are part of the NNLO corrections. For details see text.}
\end{figure}

The present experimental constraints on $HH$ production, e.g., by the CMS experiment~\cite{CMS:2022cpr} which sets an observed upper limit on the $HH$ cross section of 3.9 %(7.8) 
times the SM prediction in the $4b$ channel can then be understood as $a \in[-1.2 , 1.6  ]$, $b \in [-2.4 , 2.2 ]$ and $\kappa_3 \in [-0.82,2.94]$ at the $95\%$ confidence level (not including the impact of $a$ on the tree-level Higgs branching ratios\footnote{Obviously, the single Higgs modifiers will be dominantly constrained from single Higgs physics in a more comprehensive fit to LHC data, see~\cite{Anisha:2022ctm}.}, keeping all non-varied parameters to unity, including $\kappa_4=1$, and assuming flat QCD corrections). The lower limit on $a$ and the upper limits on $b,\kappa_3$ arise from a breakdown of perturbation theory when the inclusive cross sections cross zero. Note that this $\kappa_3$ limit is a stronger constraint compared to leading order precision $\kappa^\text{LO}_3\in [-1.03,5.81]$ (neither limit includes the impact from experimental selections which somewhat reduces the sensitivity~\cite{CMS:2022cpr}). The upper limit inferred from LO analyses alone is, therefore, currently, deeply in the strongly coupled parameter regime.\footnote{Turning to positive-definite ``squared'' NLO contributions, cf.~Fig.~\ref{fig:kfac}, the CMS result amounts to $\kappa_3  \in [-0.80, 4.62]$.}

Assuming the SM outcome prevails during the LHC's high luminosity phase, one can expect a sensitivity of $pp\to HH$ in the range of $-2.3 \lesssim \kappa^{\text{LO}}_3 \lesssim 6$ to $0 \lesssim \kappa^{\text{LO}}_3 \lesssim 2$ depending on the final state, Ref.~\cite{Cepeda:2019klc} (these extrapolations do not include weak corrections). The shortfalls of the current constraints with regards to perturbativity are self-correcting as more data becomes available towards the HL-LHC phase: the radiative corrections become controlled for $\kappa_3$ values in this parameter region, see Fig.~\ref{fig:kfac}. Triple Higgs production has not been forecast in a similar community-wide approach yet, but proof-of-principle investigations suggest very loose constraints on the quartic Higgs couplings $|\kappa_4|\lesssim 30$~\cite{Stylianou:2023xit}, which is therefore unlikely to play a significant role in more global fits beyond the $HH$ final states. From the momentum structure of Eqs.~\eqref{eq:3pt} and \eqref{eq:4pt}, it is clear that non-trivial momentum dependencies can profoundly reshape the phenomenology of these final states. In particular, owing to the singlet nature of the Higgs boson, the triple Higgs final states open up entirely new territory compared to $HH$ production (see also Ref.~\cite{Delgado:2023ynh}).\footnote{Aspects related to the special role of Eq.~\eqref{eq:2pt} have been discussed recently in Refs.~\cite{Anisha:2023ltp,Anisha:2024xxc}.} The combination of both observations leverages triple Higgs final states to add value to the HL-LHC Higgs programme even if the expectation in the SM is unpromising. We will mainly focus on 13 TeV LHC collisions, but comment on the relevance of our findings for a future FCC-hh in passing.

%%%%%%%%%%%%%%%%%%%%%%%%%%%%%%%%%%%%%%%%%%%%%%%%%%%%%
%%%%%%%%%%%%%%%%%%%%%%%%%%%%%%%%%%%%%%%%%%%%%%%%%%%%%

In the following, we will identify $a=b=1$ as these couplings will be predominantly constrained by single Higgs measurements. Note also that as we are assuming the on-shell scheme with HEFT parameters chosen to reproduce the SM for the Higgs interactions with $W,Z$ bosons (as well as SM-like fermions), the dominant decay phenomenology of the Higgs is SM-like and our results therefore generalize to the exclusive decay channels $H\to b\bar b ,\tau \tau, \gamma \gamma, WW$. 

Having a more detailed look at Fig.~\ref{fig:kfac}, we gather a qualitative estimate of the size of the radiative weak corrections within the approximation detailed above.  As can be seen, close to the SM-like choices $\kappa_3=\kappa_4=1$, the radiative corrections are moderately small $\sim 10\%$ for $HH$ and $\sim 30 \%$ for $HHH$ production (the latter results from a larger sensitivity to $\kappa_3$ as the $\kappa_4$-dependence is probed highly off-shell). These numerical NLO HEFT results for the cross sections are in concordance with the numerical results obtained in the appendix~\ref{1PIfunctions}  for the 1PI functions and indicate that the effective momentum for the virtuality $q$ value of the propagating internal Higgs boson at the LHC is rather close to the threshold value, i.e. $q \simeq 2 m_H$ and $q \simeq 3 m_H$ in $HH$ and $HHH$ production, respectively. The size of the HEFT radiative corrections in $HH$  and in $HHH$ production are in any case larger than the weak radiative corrections within the SM framework which,  according to the results for the 1PI  in the appendix,  are expected to be below $3\%$ for $HH$ production and below $9\%$ for $HHH$ production.  A main message from this Fig.~\ref{fig:kfac} is that the sensitivities to $\kappa_3$ and $\kappa_4$ change in the HEFT with respect to the LO computation due to the one-loop weak radiative corrections which include the non-linearity HEFT features and are highly sensitive to the off-shell momentum of the internal propagating Higgs bosons attached to the 1PI blob functions (see appendix~\ref{1PIfunctions} for details on these 1PI functions).
In Fig.~\ref{fig:kfac},  we also include the squared NLO contribution which is formally part of the 2-loop contribution to the Higgs self-interactions. These give a measure of strong coupling when NLO and NLO$^2$ differ significantly. The robustness of this measure ultimately requires a full 2-loop calculation which is beyond the scope of this work, but the radiative corrections become parametrically more relevant when moving further away from the SM-like choices.

%%%%%%%%%%%%%%%%%%%%%%%%%%%%%%%%%%%%%%%%%%%%%%%%%%%%%
\begin{figure}[!t]
\centering
\includegraphics[width=0.49\textwidth]{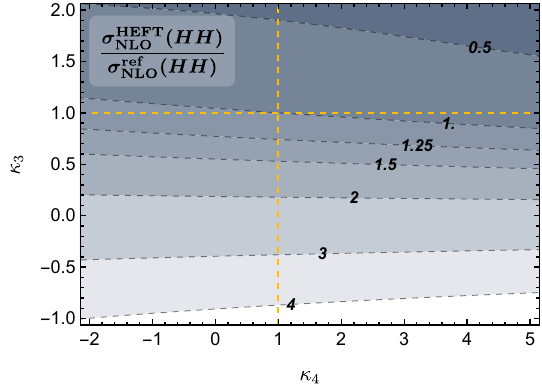}\hfill
\includegraphics[width=0.49\textwidth]{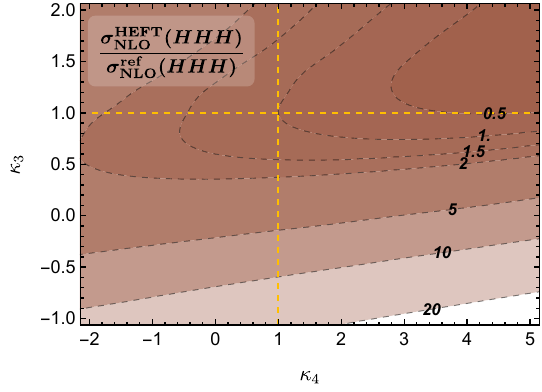}
\caption{\label{fig:2d1} Cross section contours within the NLO HEFT for $13~\text{TeV}$ $HH$ (left) and $HHH$ (right) production,  relative to the NLO reference expectation as defined in Eqs.~\eqref{eq:smxsec} and \eqref{eq:SMversusref2}. The contours are in the $(\kappa_4, \kappa_3)$ plane.}
\end{figure}
%%%%%%%%%%%%%%%%%%%%%%%%%%%%%%%%%%%%%%%%%%%%%%%%%%%%%

The radiative corrections to $pp \to HH$ further enhance the cross section for $\kappa_3<0$, which results from constructive interference of the involved triangle and box diagrams at leading order. For $\kappa_3>0$, the radiative corrections become decreasingly relevant in the region that we consider in this work $-1\lesssim \kappa_{3} \lesssim 2$, which is the region that will likely be explored during the HL-LHC phase~\cite{Cepeda:2019klc}. As $\kappa_4$ enters as a genuine higher-order effect in $pp \to HH$, the cross section dependence is relatively mild, deriving from a radiative contribution to the trilinear Higgs vertex. The $\kappa_3$ region that the LHC should explore at the high-luminosity phase, given Eq.~\eqref{eq:smxsec} can be considered relatively stable from an electroweak point of view, in particular for $HH$ production that is most relevant there.

%%%%%%%%%%%%%%%%%%%%%%%%%%%%%%%%%%%%%%%%%%%%%%%%%%%%%
\begin{figure}[!t]
\centering
\includegraphics[width=0.49\textwidth]{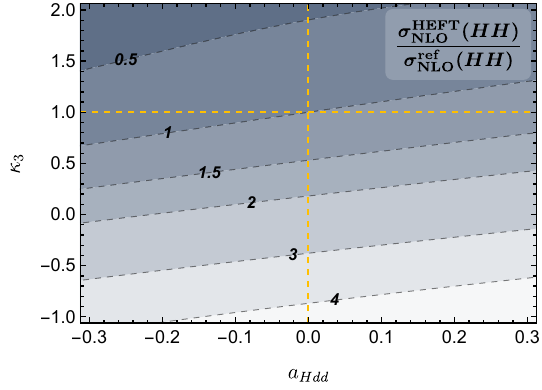}\hfill
\includegraphics[width=0.49\textwidth]{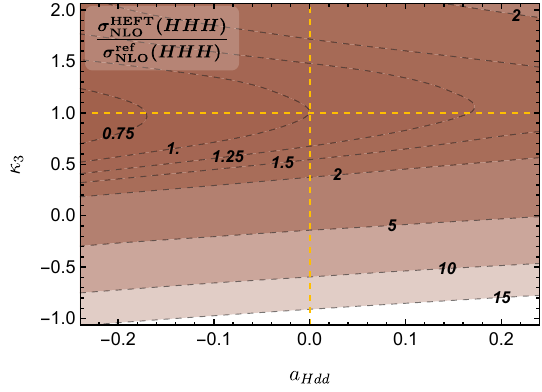}\\[0.2cm]
\includegraphics[width=0.49\textwidth]{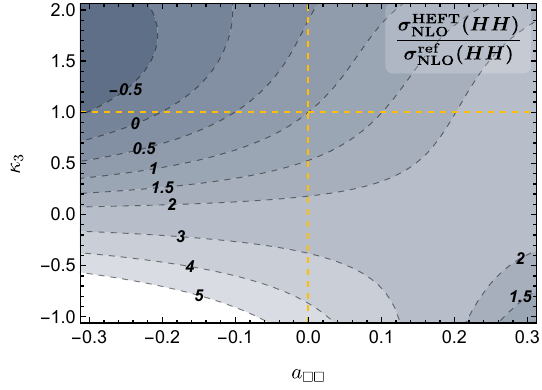}\hfill
\includegraphics[width=0.49\textwidth]{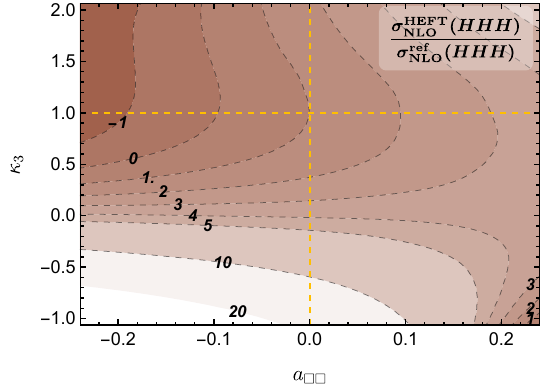}
\caption{\label{fig:2d1b} Cross section contours within the NLO HEFT for $13~\text{TeV}$ $HH$ (left) and $HHH$ (right) production,  relative to the NLO reference expectation as defined in Eqs.~\eqref{eq:smxsec} and \eqref{eq:SMversusref2}. The contours are in the $(a_{Hdd}, \kappa_3)$ plane (first row) and the $(a_{\Box \Box}, \kappa_3)$ plane (second row).}
\end{figure}
%%%%%%%%%%%%%%%%%%%%%%%%%%%%%%%%%%%%%%%%%%%%%%%%%%%%%

As $pp\to HHH$ offers richer final state kinematics, with cross sections predominantly driven by the Higgs trilinear interactions, the radiative corrections are more relevant than for $pp \to HH$. Similar to $HH$ production where the lower 3-point function destructively interferes with the box topologies, the biggest share of the cross section arises from the pentagon contributions of Fig.~\ref{fig:Feyn}(c), which are destructively offset against the triangle and box contributions, which interfere constructively. $\kappa_3<0$, therefore, leads to an enhancement of the cross section whilst the $\kappa_4$ dependence is relatively flat due to the $s$-channel suppression. The wider kinematical phase space compared to $HH$ production, however, enables cancellations between the different effective topologies in different phase space regions. Overall this leads to a reduced cross section via a softer distribution for $m_{HHH}\gtrsim 3m_t$, when including radiative corrections with $\kappa_3=\kappa_4=1$. We can therefore expect that non-standard top interactions that are not discussed in this work are likely to be visible in this channel, too.

%%%%%%%%%%%%%%%%%%%%%%%%%%%%%%%%%%%%%%%%%%%%%%%%%%%%%
\begin{figure}[!t]
\centering
\includegraphics[width=0.49\textwidth]{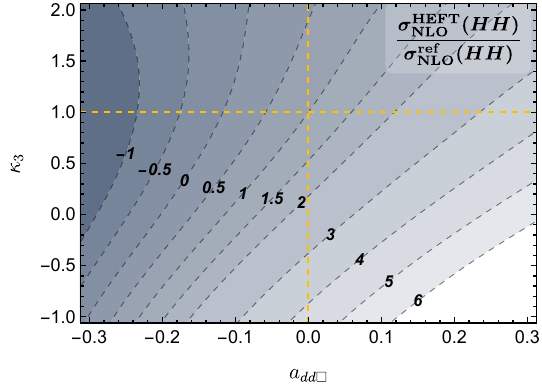}\hfill
\includegraphics[width=0.49\textwidth]{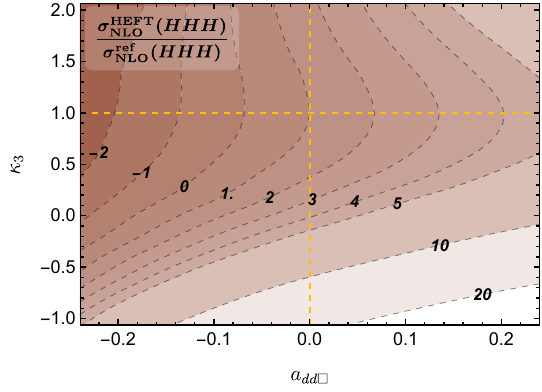}\\[0.2cm]
\includegraphics[width=0.49\textwidth]{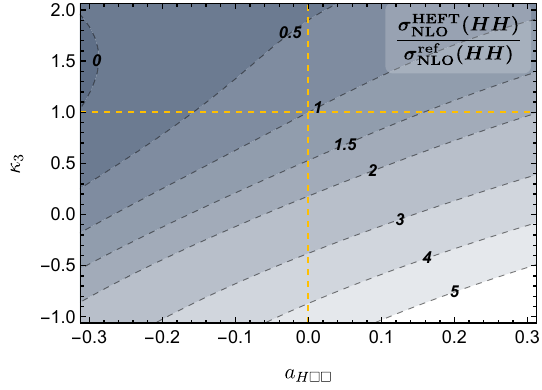}\hfill
\includegraphics[width=0.49\textwidth]{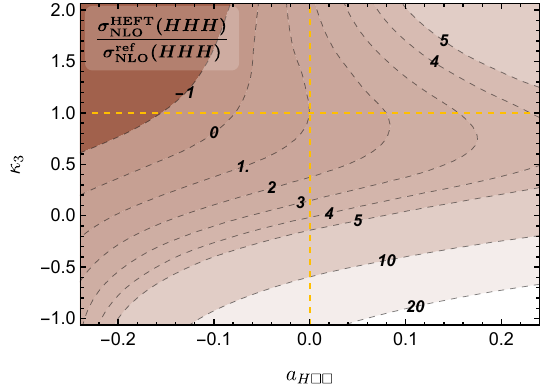}
\caption{\label{fig:2d2} Cross section contours within the NLO HEFT for $13~\text{TeV}$ $HH$ (left) and $HHH$ (right) production,  relative to the NLO reference expectation as defined in Eqs.~\eqref{eq:smxsec} and \eqref{eq:SMversusref2}. The contours are in the $(a_{dd\Box}, \kappa_3)$ plane (first row) and the $(a_{H\Box \Box}, \kappa_3)$ plane (second row).}
\end{figure}
%%%%%%%%%%%%%%%%%%%%%%%%%%%%%%%%%%%%%%%%%%%%%%%%%%%%%

\begin{figure}[!t]
\centering
\includegraphics[width=0.49\textwidth]{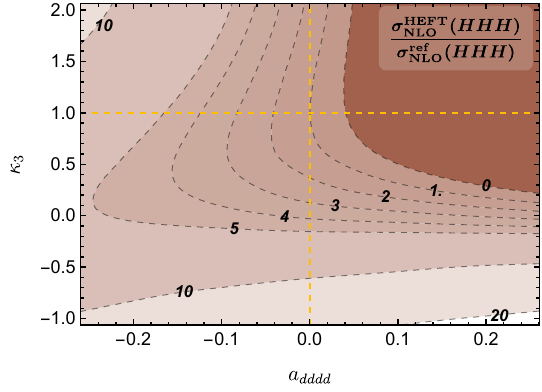}\hfill
\includegraphics[width=0.49\textwidth]{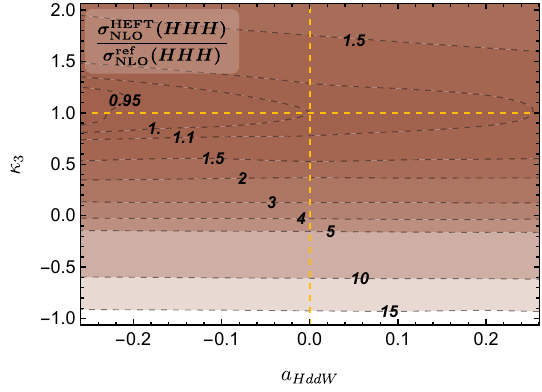}\\[0.2cm]
\includegraphics[width=0.49\textwidth]{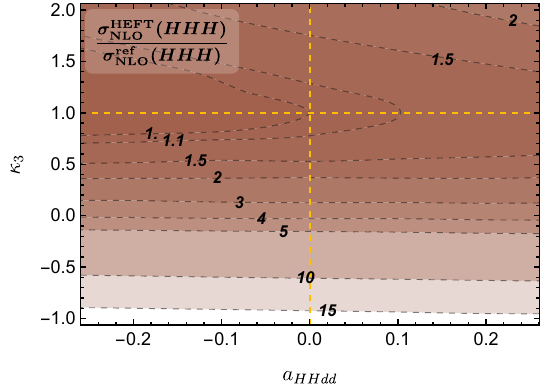}\hfill
\caption{\label{fig:2d3}  Cross section contours within the NLO HEFT for $13~\text{TeV}$  $HHH$  production,  relative to the NLO reference expectation as defined in Eqs.~\eqref{eq:smxsec} and \eqref{eq:SMversusref2}.  The contours are in the $(a_{dddd}, \kappa_3)$ plane (upper left plot),  the $(a_{HddW}, \kappa_3)$ plane (upper right plot) and the $(a_{HHdd}, \kappa_3)$ plane (lower plot).}
\end{figure}

From the above discussion, it becomes clear that visible enhancements via $\kappa_4$ at the LHC in $HHH$ production, in particular when considering SM parameter choices of the HEFT coupling space, are difficult to obtain (see also Refs.~\cite{Papaefstathiou:2023uum,Stylianou:2023xit}). Limiting ourselves to ${\cal{O}}(p^2)$ operators, this process is predominantly driven by modifications of the trilinear Higgs coupling, which is much better accessible with $HH$ production, cf.~Fig.~\ref{fig:2d1}. New opportunities arise from the ${\cal{O}}(p^4)$ interactions. As $\kappa_3$ is a relevant parameter given the above discussion, we show pairwise correlations of the relevant interactions, as a ratio against the SM parameter choice $\kappa_3=\kappa_4=a=b=1$ (with top quarks taken SM-like) in Figs.~\ref{fig:2d1b}, \ref{fig:2d2}, and \ref{fig:2d3}, where we focus on cross sections expanded according to Eq.~\eqref{eq:ampexp}.\footnote{Cross sections are therefore not positive definite and negative cross sections are typically understood as parameter choices giving rise to strong coupling as the $S$ matrix remains unitary.} More concretely, in the following, we refer to
\begin{equation}
\label{eq:smxsec}
\sigma_\text{NLO}^{\text{ref}}=\sigma^{\text{HEFT}}_{\text{NLO}}(\kappa_3,\kappa_4,a,b=1,a_i=0)~\text{for all remaining $i$ in Eq.~\eqref{eq:lag},}
\end{equation}
as the reference value (see, e.g.~\cite{Englert:2023uug} for a discussion on how this choice can be related to the SM through field redefinitions). When scanning over the HEFT parameters we will always retain a choice $a=b=1$.   

It should be noticed that this NLO reference cross section value in Eq. ~\eqref{eq:smxsec} is not far numerically from the SM-LO prediction. For the LHC with 13 TeV collisions we observe
\begin{eqnarray}
%\label{eq:SMversusref}
&\sigma^\text{SM}_\text{LO} (HH)=\sigma_\text{LO}^\text{ref} (HH) = 17.40 \, {\text{fb}}\, ; \sigma^\text{SM}_\text{LO}  (HHH) =  \sigma_\text{LO}^\text{ref} (HHH)= 0.041 \,  {\text{fb}}\,  \label{eq:SMversusref1}\\
&\sigma_\text{NLO}^\text{ref} (HH)  =  18.45 \,  {\text{fb}}\, ;  \sigma_\text{NLO}^\text{ref} (HHH) =  0.029 \,  {\text{fb}}\,. \label{eq:SMversusref2}
\end{eqnarray}
Notice again that, according to the results for the 1PI insertions in the appendix, the SM NLO corrections are expected to be very small (below $3\%$ for $HH$ production and below $9\%$ for $HHH$ production) and,  in consequence, the SM LO  provide quite stable SM rates. Then, for illustrative purposes,  the forthcoming results for the ratios $\sigma_\text{ NLO}^\text{ HEFT}(HH)/\sigma_\text{NLO}^\text{ref}(HH)$ and $\sigma_\text{NLO}^\text{HEFT}(HHH)/\sigma_\text{NLO}^\text{ref}(HHH)$ can be easily translated into results for $\sigma_\text{NLO}^\text{HEFT}(HH)/\sigma_\text{LO}^\text{SM}(HH)$ and $\sigma_\text{NLO}^\text{HEFT}(HHH)/\sigma_\text{LO}^\text{SM}(HHH)$ by simply rescaling them with factors 1.06 and 0.71,  respectively.   The important feature is that any of these ratios will provide sensitivity to the HEFT coefficients via NLO corrections within the HEFT. As explained fully in the appendix,  these NLO corrections are of two types,  one-loop corrections and corrections from the  ${\cal{O}}(p^4)$ coefficients.  These two corrections affect the momenta dependence of the scattering amplitudes and result in notable departures of the cross section predictions with respect to the SM predictions.  In particular,  the change in the dependence with the virtuality $q$ of the internal propagating Higgs bosons plays a very relevant role in these departures and is basically due to the non-linearity in the boson interactions within the HEFT (see the appendix for more details). 
More specifically, Figs.~\ref{fig:2d1b} and \ref{fig:2d2} display the interactions that affect,  via radiative corrections and NLO operators, both $HH$ and $HHH$ production.   Interactions such as $\sim a_{Hdd}, a_{ddW}, a_{ddZ}$, for the on-shell final state Higgs momenta are qualitatively similar to $\kappa_3$ modifications as the threshold of double and triple Higgs production dominates the cross section.

 In these cases, the cross section modifications largely follow the ${\cal{O}}(p^2)$ paradigm, see Fig.~\ref{fig:2d1b} (showing a significant correlation in $HH$). Non-trivial momentum dependencies, such as $a_{\Box\Box}$, create less standard scalings through the $q^4$ enhancement of the off-shell propagator that is accessed in both $HH$ and $HHH$ production (e.g. in $HH$ production a decrease in the 3-point contribution can be compensated by an enhancement in the two-point function giving rise to the characteristic contours for $a_{\Box\Box}$ vs. $\kappa_3$).

%%%%%%%%%%%%%%%%%%%%%%%%%%%%%%%%%%%%%%%%%%%%%%%%%%%%%
%%%%%%%%%%%%%%%%%%%%%%%%%%%%%%%%%%%%%%%%%%%%%%%%%%%%%

As becomes clear for these interactions already, non-trivial momentum dependencies that are attributed to the Higgs boson's singlet character in HEFT, can indeed alter the cross sections significantly leading to a different contour structure as different phase space regions are modified in different and characteristic ways (e.g. by comparing Fig.~\ref{fig:2d2} against the momentum behaviour tabled in Eqs.~\eqref{eq:2pt}-\eqref{eq:4pt}. When turning to interactions that exclusively impact the four-point interactions, Fig.~\ref{fig:2d3}, these are typically suppressed due to the off-shellness with which they are probed as contributions to triple Higgs production. The notable exception here is $a_{dddd}$ which signifies a dramatic momentum-dependent departure from the SM thus leading to cross section enhancements, particularly when $HH$ measurements indicate $\kappa_3\simeq 1$ in the future. Our findings therefore reflect qualitative the power counting of HEFT so that $\kappa_3$ as a chiral dimension 2 operator that does not suffer from the kinematical drawbacks of $\kappa_4$ shapes the $HHH$ cross sections, yet with significant complementarity compared to $HH$. 

%%%%%%%%%%%%%%%%%%%%%%%%%%%%%%%%%%%%%%%%%%%%%%%%%%%%%
\begin{figure}[!t]
\centering
\parbox{0.49\textwidth}{{\includegraphics[width=0.49\textwidth]{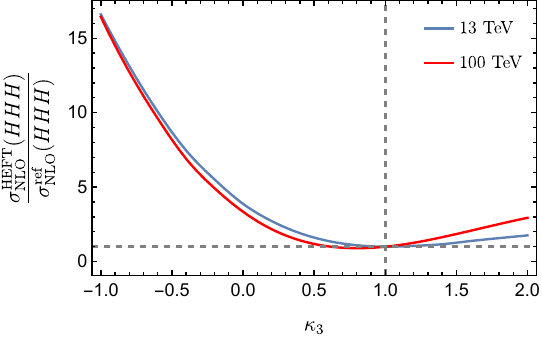}}}\hfill
\parbox{0.49\textwidth}{{\includegraphics[width=0.49\textwidth]{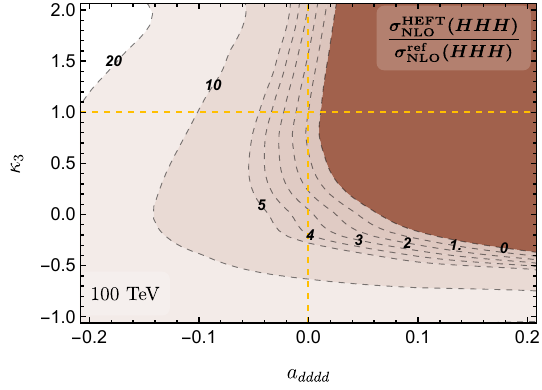}}}
\caption{\label{fig:100tev}  (Left plot) Cross section predictions within the NLO HEFT for $100~\text{TeV}$ $HHH$ production and comparison with $13~\text{TeV}$. These predictions are again relative to the NLO reference expectation of Eq. ~\eqref{eq:smxsec}.  Corrections are quantitatively similar to the LHC case, generalising our 13 TeV findings also to collisions at higher energy.  (Right plot) Contours in the $(a_{dddd}, \kappa_3)$ plane. }
\end{figure}
%%%%%%%%%%%%%%%%%%%%%%%%%%%%%%%%%%%%%%%%%%%%%%%%%%%%%

Of course, all of our discussion needs to be viewed against the backdrop of a small $HHH$ baseline rate. Intrinsically this makes order-1 modifications not directly phenomenologically relevant, in particular as we can expect a realistic reconstruction in a busy environment to be experimentally challenging. That said, the HL-LHC is likely to test the $\kappa_3 \gtrsim -0.5$; this is a parameter region where a significant, order-of-magnitude departure of the $HHH$ production is possible provided that the Higgs potential reflects a sizeable amount of non-linearity. Such enhancements could be probable at the HL-LHC and future colliders and might constitute a strong motivation to pursue this final state in whilst closing in on the HL-LHC $HH$ production sensitivity target.

Looking towards the more distant future, in Fig.~\ref{fig:100tev} we show representative distributions of cross section modifications at a 100 TeV FCC-hh. Our findings here are qualitatively similar to those for 13 TeV collisions, of course, at a significantly increased SM production rate. In particular, the momentum dependencies are therefore probed with a higher expected sensitivity. And as 100 TeV collisions probe the tails of $HHH$ production much more efficiently than the LHC, we see larger enhancements for parameter choices, that populate the phase space regions of large momentum transfers, e.g. $a_{dddd}$ which is shown alongside the $\kappa_3$ LHC/FCC-hh comparison in Fig.~\ref{fig:100tev}. These observations tie into the existing results of 100~TeV triple Higgs production presented in various model contexts, see Refs.~\cite{Chen:2015gva,Fuks:2015hna,Papaefstathiou:2015paa,Fuks:2017zkg,Kilian:2017nio,Papaefstathiou:2019ofh,Papaefstathiou:2023uum}.

%%%%%%%%%%%%%%%%%%%%%%%%%%%%%%%%%%%%%%%%%%%%%%%%%%%%%
\section{Summary and Conclusions}
\label{sec:conc}
%%%%%%%%%%%%%%%%%%%%%%%%%%%%%%%%%%%%%%%%%%%%%%%%%%%%%
The production of multiple Higgs bosons at the LHC and beyond is a powerful tool for discerning the nature of the electroweak scale by directly investigating the symmetry-breaking dynamics. This comes at the price of relatively small (in case of $HH$ production) and challenging (in case of $HHH$ production) cross sections in the SM. However, BSM modifications of the Higgs potential at the TeV scale can lead to significant alterations of cross sections. In particular, when underlying parameters of the Higgs potential are very far from the SM point, cross sections can be significantly enhanced, making radiative corrections exceedingly relevant. In this work, adopting the framework of Higgs Effective Field Theory, we have investigated {\emph{all}} dominant bosonic modifications of the dominant production modes $gg\to HH(H)$ including one-loop radiative corrections in the HEFT framework up to ${\cal{O}}(p^4)$ in the momentum expansion. We find that the power-counting expected from the HEFT construction is largely reflected in the cross section enhancements observed in the $HH(H)$ rates. This further motivates $HH$ physics whilst moving towards the HL-LHC phase. If the TeV scale contains aspects of non-linearity, they will show up in pronounced ways in di-Higgs production. Owing to its small rate in the SM, triple Higgs production remains challenging also when turning to HEFT. However, parameter regions exist where triple Higgs production can provide complementary information, at a very large enhancement, for which $HH$ production would show an anomaly in the range of its HL-LHC sensitivity extrapolation. Such a situation could then lead to a direct discovery of BSM aspects of the TeV scale, which could be probed in much more detail at future hadron-hadron machines.

%%%%%%%%%%%%%%%%%%%%%%%%%%%%%%%%%%%%%%%%%%%%%%%%%%%%%
\section*{Acknowledgments}
%%%%%%%%%%%%%%%%%%%%%%%%%%%%%%%%%%%%%%%%%%%%%%%%%%%%%
We dedicate this work to Peter Higgs, whose work underpins all of ours.\\

\noindent A. is funded by the Leverhulme Trust under Research Project Grant RPG-2021-031. 
D.D. and M.J.H. acknowledge financial support from the grant IFT Centro de Excelencia Severo Ochoa CEX2020-001007-S funded by MCIN/AEI/10.13039/ 501100011033, from the Spanish ``Agencia Estatal de Investigaci\'on'' (AEI) and the EU ``Fondo Europeo de Desarrollo Regional'' (FEDER) through the project PID2019-108892RB-I00 funded by MCIN/AEI/ 10.13039/501100011033, and from the European Union’s Horizon 2020 research and innovation programme under the Marie Sklodowska-Curie Grant agreement No. 860881-HIDDeN. D. D. and M. J. H. also acknowledge partial financial support by the Spanish Research Agency (Agencia Estatal de Investigaci\'on) through the Grant PID2022-137127NB-I00 funded by MCIN/AEI/10.13039/501100011033/FEDER, UE. 
C.E. is supported by the UK Science and Technology Facilities Council (STFC) under grant ST/X000605/1 and the Leverhulme Trust under Research Project Grant RPG-2021-031.
The work of R.A.M. is supported by CONICET and ANPCyT under project PICT-2021-00374.

%%%%%%%%%%%%%%%%%%%%%%%%%%%%%%%%%%%%%%%%%%%%%%
\section*{Appendix}
\appendix
%%%%%%%%%
\section{Details of the 1PI Higgs functions}
\label{1PIfunctions}
%\textcolor{red}{MH: this annex could be here or could be a section or subsection, as you prefer}
%%%%%%%%%
%%%%%%%%%%%%%%%%%%%%%%%%%%%%%%%%%%%%%%%%%%%%%%%%%%%%%
\begin{figure}[!t]
\centering
\includegraphics[width=1\textwidth]{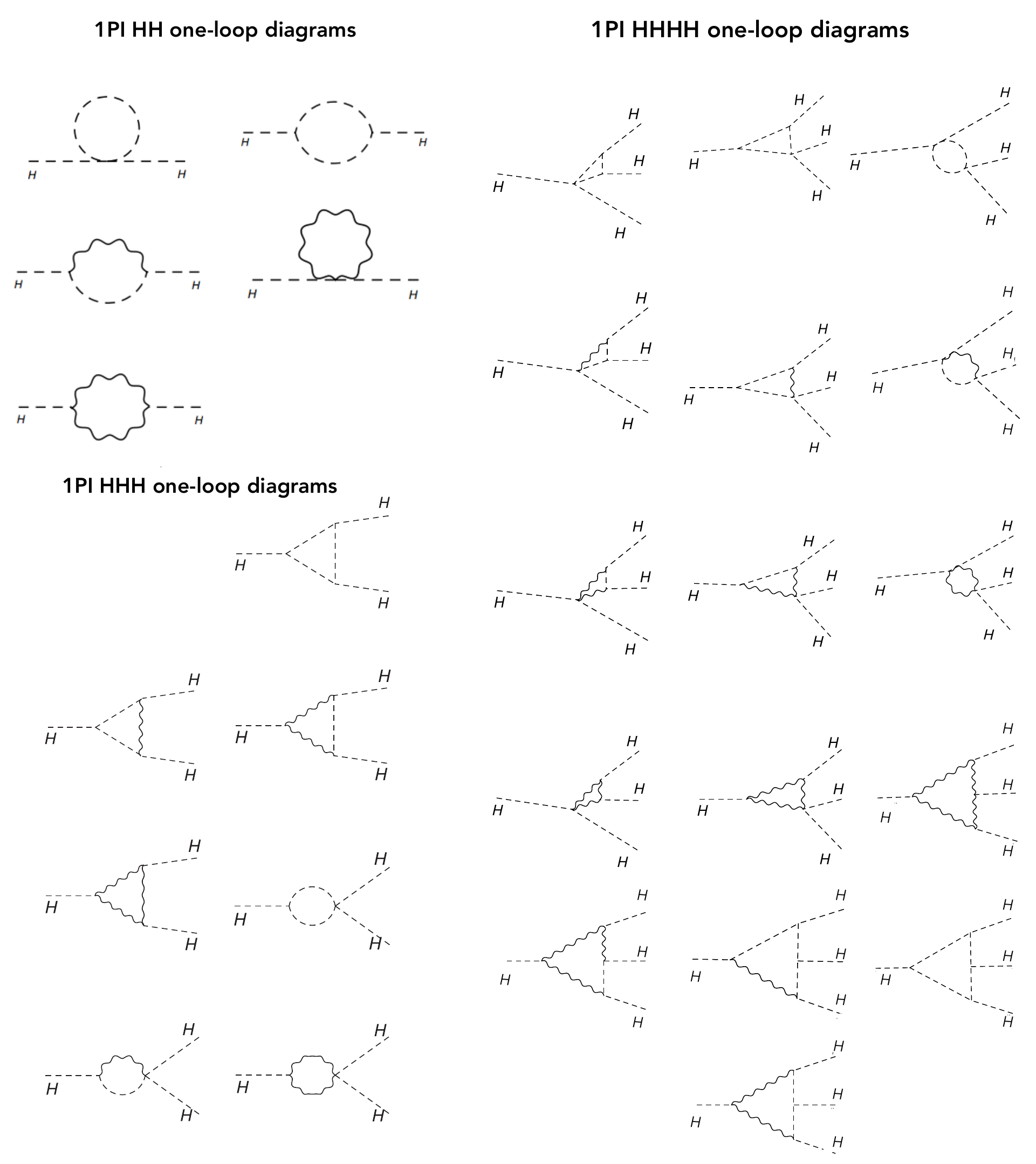}
\caption{1PI bosonic loops.  Here wavy lines denote generically gauge bosons,  $W$ and/or $Z$,  and dashed lines denote generically scalar bosons,  $H$ and/or GBs $\pi$.}
\label{1PIloops}
\end{figure}
%%%%%%%%%%%%%%%%%%%%%%%%%%%%%%%%%%%%%%%%%%%%%%%%%%%%
Here we analyse the 1PI Higgs functions involved in this work.  The results of the renormalized 1PI functions are presented schematically in  Eqs.~\eqref{eq:2pt}, \eqref {eq:3pt},  and \eqref {eq:4pt} for the cases of 2, 3 and 4 external Higgs legs, respectively. To get the analytical and numerical results we have implemented the HEFT model with {\tt{FeynRules}}~\cite{Christensen:2008py,Alloul:2013bka} and performed the main calculations with {\tt{FormCalc}} and {\tt{LoopTools}}~\cite{Hahn:2000kx,Hahn:1998yk,Hahn:2000jm,Hahn:2006qw}.  We use here the Feynman-'t Hooft gauge and regularize the integrals with dimensional regularization in $D=4-\epsilon$ dimensions.  For the renormalization prescription we apply the on-shell renormalization conditions for the $H$, $Z$ and $W$ bosons and the HEFT coefficients are renormalized in the \MSb~scheme. The set of relevant HEFT Feynman rules for this computation in covariant $R_\xi$ gauges as well as the details of the renormalization procedure can be found in~\cite{Herrero:2020dtv,Herrero:2021iqt,Herrero:2022krh}.  For comparison,  we have also computed in parallel the renormalized 1PI functions within the SM framework to one-loop and with on-shell conditions for the SM $H$, $Z$ and $W$ bosons. In all cases, the tadpole 1PI H function has been renormalized to zero value. 

The NLO HEFT results for the renormalized self-energy $\Sigma_{HH}$ are given in~\eqref{eq:2pt} and receive contributions from loop diagrams,  counterterms and from a single NLO coefficient $a_{\Box \Box}$. The bosonic loop diagrams are collected in Fig.~\ref{1PIloops}.  The NLO HEFT results for the renormalized $HHH$ and $HHHH$  1PI functions in equations~\eqref{eq:3pt} and~\eqref{eq:4pt} receive contributions from different origins:  1) from the LO Lagrangian terms,  which depend on the LO coefficients ($\kappa_3$ in $HHH$ and $\kappa_4$ in $HHHH$),  2) from the NLO Lagrangian terms,  which depend on the NLO coefficients  (generically called $a_i$'s),   3) from the one-loop diagrams collected in Fig.~\ref{1PIloops} (involving the LO coefficients $a$, $b$,  $\kappa_3$ and $\kappa_4$ in the vertices),  and 4) from counterterms,  which are generated from both the LO Lagrangian (in this case leading to the terms involving $\delta \kappa_3$,  $\delta \kappa_4$,  $\delta\Zf_H$ ,  $\delta \mh^2$,  $\delta \vev/\vev$,  and $\delta Z_\pi$), and from the NLO Lagrangian (generically called $\delta a_i$'s).  These latter are typically generated by the shift $a_i \to a_i+\delta a_i$ and are needed in the HEFT to remove the extra divergences appearing from the loop diagrams in addition to the divergences removed by the previous counterterms in the LO Lagrangian. These four contributions can be written generically as follows:
\be
\greenfR^{\rm NLO}=\Gamma^{\rm LO}+\Gamma^{a_i}+\greenfL +\greenfC \,,
\label{split1} 
\ee
where,  
\be
\Gamma^{\rm LO}_{HHH} = -3 \kappa_3\frac{\mh^2}{\vev}  \,,  \quad
\Gamma^{\rm LO}_{HHHH}=-3\kappa_4\frac{\mh^2}{\vev^2} \,,
\label{LO1PI}
\ee
and the NLO coefficients that enter in the $\Gamma^{a_i}$'s are:  $a_{H\Box\Box}$,  $a_{dd\Box}$,  $a_{Hdd}$,   $a_{ddW}$ and $a_{ddZ}$ for $HHH$,  and $a_{dddd}$,  $a_{HH\Box\Box}$,  $a_{Hdd\Box}$,  $a_{HHdd}$, $a_{HddW}$ and $a_{HddZ}$ for $HHHH$ (see Tab.~\ref{tab:operators}). In addition $a_{\Box\Box}$ enters in both $HHH$ and $HHHH$ via the finite contributions to $\delta \Zf_H$ and $\delta \mh^2$ in the on-shell scheme.   Specifically, 
\begin{equation}
\delta Z_H=\Sigma_{HH}^{\rm ' Loop}(q^2=m_H^2)- \frac{4m_H^2}{v^2} a_{\Box \Box} \, ,  \quad
\delta m_H^2 = -\Sigma_{HH}^{\rm Loop}(q^2=m_H^2) - \frac{2m_H^4}{v^2} a_{\Box \Box} \, . 
\end{equation}

The renormalized 1PI functions for $HH$, $HHH$ and $HHHH$ have been checked to be finite for all values of the external momenta. For this finiteness check  just the ${\cal O}(\Delta_\epsilon)$ pieces (named $\deltaCT$) of the counterterms and the HEFT coefficients in Eqs.~\eqref{eq:2pt}, \eqref {eq:3pt},  and \eqref{eq:4pt} are needed.  We include them below for completeness, 
\bear
&&\deltaCT\Zf_H=\frac{\div}{16\pi^2}\frac{3a^2}{\vev^2}(2\mw^2+\mz^2)\,,  \qquad \div=\frac{2}{\epsilon}-\gamma_E+\log(4\pi) \,,
\nn\\
&&\deltaCT\mh^2=\frac{\div}{16\pi^2}\frac{3}{2\vev^2}((3\kappa_3^2+\kappa_4)\mh^4-2a^2\mh^2(2\mw^2+\mz^2)+(4a^2+2b)(2\mw^4+\mz^4))\,,  \nn\\
&&\deltaCT\Zf_\pi=-\frac{\div}{16\pi^2}\frac{1}{\vev^2}((a^2-b)\mh^2-(3a^2+5/3)\mw^2-5/3\mz^2)\,,  \nn\\
&&\deltaCT\vev/\vev=\frac{\div}{16\pi^2}\frac{2(\mw^2+\mz^2)}{3\vev^2}\,, \nn\\
&&\deltaCT\kappa_3 = -\frac{\Delta_\epsilon}{16\pi^2}\frac{1}{2\mh^2\vev^2}\left(\kappa_3(a^2-b+9\kappa_3^2-6\kappa_4)\mh^4-3(1-a^2)\kappa_3\mh^2(\mw^2+\mz^2)\right.  \nn\\
&&\left. \hspace{15mm} +6(-2ab+2a^2\kappa_3+b\kappa_3)(2\mw^4+\mz^4)\right)\,, \nn\\
&& \deltaCT\kappa_4 = -\frac{\Delta_\epsilon}{16\pi^2}\frac{1}{2\mh^2\vev^2}\left(\kappa_4(2a^2-2b+9\kappa_3^2-6\kappa_4)\mh^4-6(1-a^2)\kappa_4\mh^2(\mw^2+\mz^2)\right.  \nn\\
&&\left. \hspace{15mm} +6(-2b^2+2a^2\kappa_4+b\kappa_4)(2\mw^4+\mz^4)\right)\,, \nn\\
&&\deltaCT a_{\Box\Box}=-\frac{\Delta_\epsilon}{16\pi^2}\frac{3a^2}{4} \,,\qquad \deltaCT a_{H\Box\Box}=\frac{\Delta_\epsilon}{16\pi^2}\frac{3a(2a^2-b)}{2}\,,  \nn\\
&&\deltaCT a_{dd\Box}=\frac{\Delta_\epsilon}{16\pi^2}\frac{3a(a^2-b)}{2}\,,\quad \deltaCT a_{Hdd}=0\,,\quad \deltaCT a_{ddW}/2=\deltaCT a_{ddZ}=-\frac{\Delta_\epsilon}{16\pi^2}3a(a^2-b)\,,  \nn\\
&&\deltaCT a_{dddd} = -\frac{\Delta_\epsilon}{16\pi^2}\frac{3(a^2-b)^2}{4}\,,\quad \deltaCT a_{HH\Box\Box} = -\frac{\Delta_\epsilon}{16\pi^2}\frac{3(12a^4-10a^2b+b^2)}{4}\,,  \nn\\
&&\deltaCT a_{Hdd\Box} = -\frac{\Delta_\epsilon}{16\pi^2}\frac{3(6a^4-7a^2b+b^2)}{2}\,,\quad 
\deltaCT a_{HHdd} = 0\,,  \nn\\ 
&&\deltaCT a_{HddW} = \frac{\Delta_\epsilon}{16\pi^2}3(4a^4-5a^2b+b^2)\,,\quad \deltaCT a_{HddZ} = \frac{\Delta_\epsilon}{16\pi^2}\frac{3(4a^4-5a^2b+b^2)}{2}\,.
\label{divergences}
\eear
The above results for  $\deltaCT\kappa_4$,  $\deltaCT a_{dddd}$,  $\deltaCT a_{HH\Box\Box}$,  
$\deltaCT a_{Hdd\Box}$,  $\deltaCT a_{HHdd}$,  $\deltaCT a_{HddW}$ and $\deltaCT a_{HddZ}$ 
are new in the literature~\footnote{The others are extracted from \cite{Herrero:2022krh}.  Notice that we have included here the last line in  $\deltaCT\kappa_3$  that was droped in the edited version of~\cite{Herrero:2022krh}. Partial checks of the above results for the HEFT divergences have been done with the previous literature. 
In Ref.~\cite{Gavela:2014uta}, the 1PI functions were renormalized to one-loop for external off-shell momenta considering the pure scalar theory (i.e. no gauge bosons included) and with massless GBs.  We find agreement in the divergences of the $a_i$s of the scalar sector,  concretely in  $a_{Hdd\Box}$ and $a_{HH\Box\Box}$ (which are $a_{\Delta H}$ and $b_{\Box H}$ in their notation).  In Ref.~\cite{Delgado:2013hxa},  the divergence of the $a_{dddd}$ coefficient ($\gamma$ in that reference) was derived within the pure scalar theory (i.e. only scalar loops) from the one-loop renormalization of $HH\to HH$ scattering.  We also find agreement with this divergence by doing the corresponding simplifications. In reference ~\cite{Asiain:2021lch},  all kind of bosonic loops (scalar and gauge bosons,  but using the Equivalence Theorem and the isospin limit with $\mw=\mz$) were computed for different scattering amplitudes and the renormalization of the trilinear and quartic Higgs couplings were derived.  Our results for $\deltaCT \kappa_3$ and $\deltaCT\kappa_4$  are in agreement with this reference after using the equation of motion of the Higgs field and simplifying our results by taking the isospin limit,  i.e. setting $\mw=\mz$.}, as far as we know.

Notice that from these divergences in~\eqref{divergences} one can easily derive the running equations for the involved HEFT coefficients,  $c_i$ (concretely $\kappa_3$,  $\kappa_4$ and the $a_i$'s):
\bear
c_i(\mu)&=&c_i(\mu')+\frac{1}{16\pi^2}\gamma_{c_i}\log\left(\frac{\mu^2}{\mu'^2}\right) \,,  \quad
\deltaCT c_i =\frac{\Delta_\epsilon}{16\pi^2}\gamma_{c_i}
\label{RGE1}
\eear
In particular, for the simplest choice of $a=b=1$ which will be the one taken in our numerical analysis in this work, there are just a few coefficients involved here that run,  specifically $\kappa_3$,  $\kappa_4$,  $a_{\Box \Box}$,  $a_{H\Box \Box}$, and $a_{HH\Box\Box}$,  with corresponding anomalous dimensions:
\bear
\gamma_{\kappa_3}&=& -\frac{1}{2\mh^2\vev^2}\left(\kappa_3(9\kappa_3^2-6\kappa_4)\mh^4 +6(3\kappa_3-2)(2\mw^4+\mz^4)\right) \nn \\
%-\frac{1}{2\vev^2}\left(\kappa_3(9\kappa_3^2-6\kappa_4)\mh^2\right)
\gamma_{\kappa_4}&=& -\frac{1}{2\mh^2\vev^2}\left(\kappa_4(9\kappa_3^2-6\kappa_4)\mh^4 +6(3\kappa_4-2)(2\mw^4+\mz^4)\right) \nn \\
%-\frac{1}{2\mh^2 v^2}2 \left(  k_4 \left(\mh^4+3\mh^2 \left(\mw^2+\mz^2\right)+6 \left(2 \mw^4+\mz^4\right)\right)-12  \left(2 \mw^4+\mz^4\right) \right. \nn\\
%&&\left. \hspace{15mm}-2  k_4 \left(\mh^4-3 \left(2 \mw^4+\mz^4\right)\right)+9 k_3^2 k_4 \mh^4-6 k_4 \mh^2 \left(k_4 \mh^2+\mw^2+\mz^2\right) \right)\,, \nn\\
\gamma_{a_{\Box\Box}}&=&-\frac{3}{4} \,,\quad
\gamma_{a_{H\Box\Box}}=\frac{3}{2} \,,\quad
\gamma_{a_{HH\Box\Box}}=-\frac{9}{4} \,.
\label{anom-dim}
\eear

Finally, in order to analyse numerically the size of the radiative corrections in these renormalized 1PI functions, it is illustrative to also split the previous results  in an alternative way, separating the contributions into tree-level ones and one-loop ones of orders $\mO(\hbar^0)$ and $\mO(\hbar^1)$, respectively:
\be
\greenfR^{\rm NLO} =\Gamma^{\rm tree}+ \Delta\Gamma^{\rm 1-loop} \,.
\label{split2a} 
\ee
where
\be
\Gamma^{\rm tree}= \Gamma^{\rm LO}+\Gamma^{\rm a_i} \,; \qquad  \Delta\Gamma^{\rm 1-loop}=\greenfL +\greenfC \,.
\label{split2b} 
\ee
 Notice that 'LO' and 'tree' only coincide if all the involved $a_i$'s of the NLO Lagrangian are assumed to be vanishing.  On the other hand,  in the self-energy case, the LO contribution is absent and there are just contributions from $\Sigma^{a_{\Box \Box}}$ and from $ \Delta\Sigma^{\rm 1-loop}$. 
 
\begin{figure}[!t]
%\centering
\includegraphics[width=0.5\textwidth]{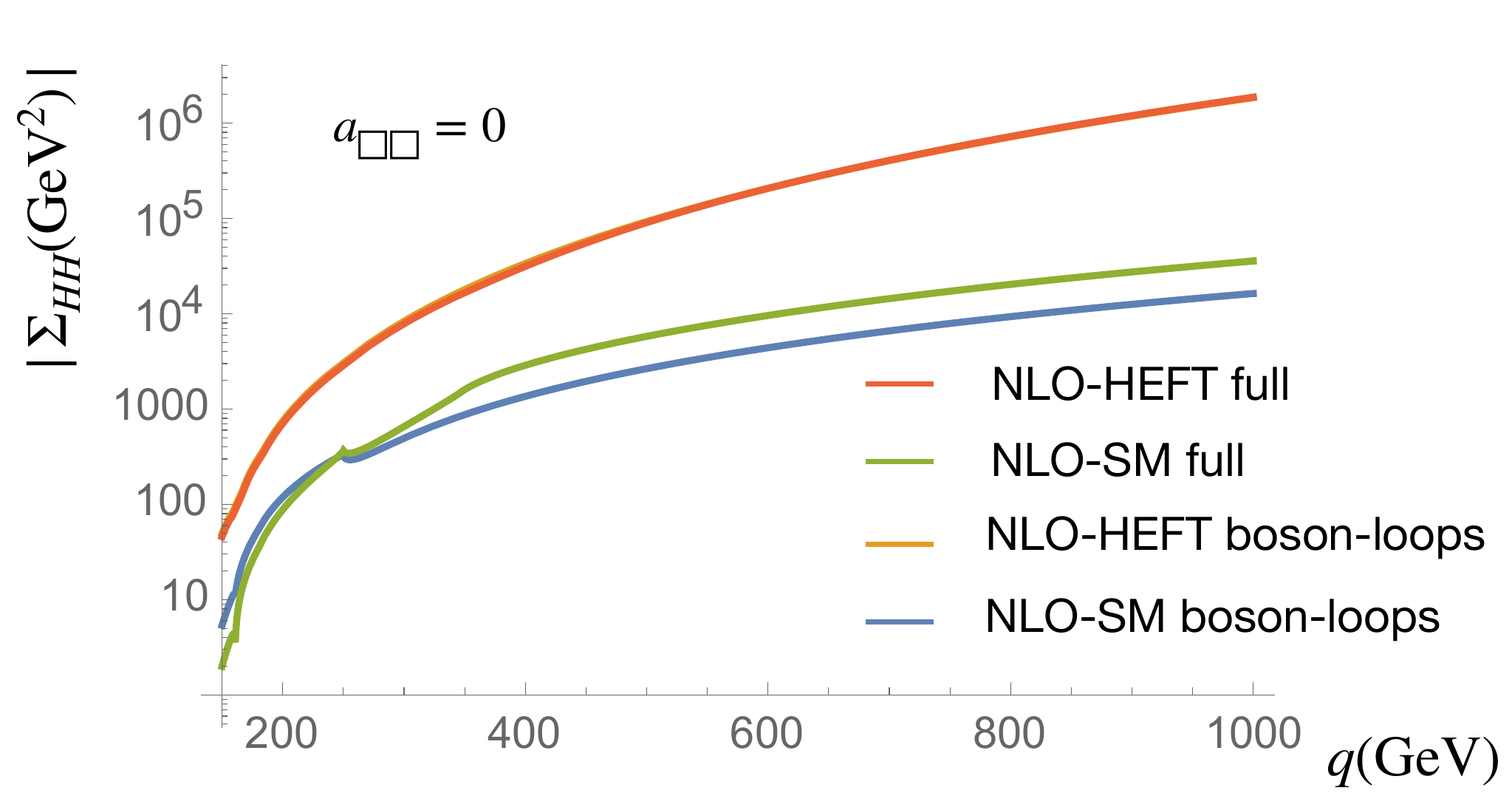}
\includegraphics[width=0.5\textwidth]{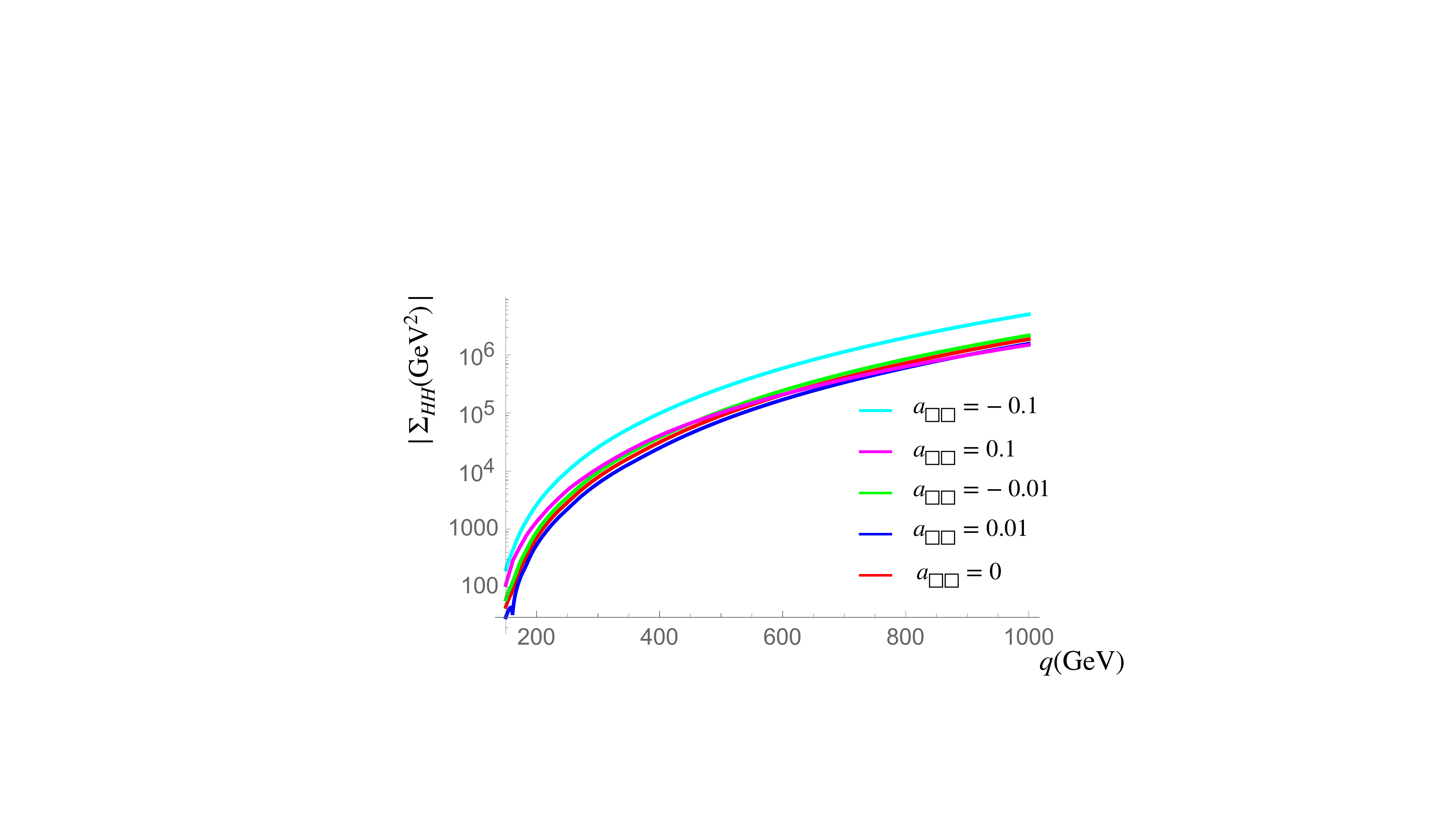}
\caption{In the left plot, the predictions for the self-energy within the NLO HEFT as a function of the off-shell momentum $q$ are shown for $a_{\Box \Box}=0$.  Both the full (boson+top loops) and the separate contributions from boson loops are displayed.  The corresponding SM predictions are also included for comparison. In the right plot, the NLO HEFT predictions for various $a_{\Box \Box}$ values are shown.}
\label{HH-q}
\end{figure}

\begin{figure}[!t]
%\centering
\includegraphics[width=0.52\textwidth]{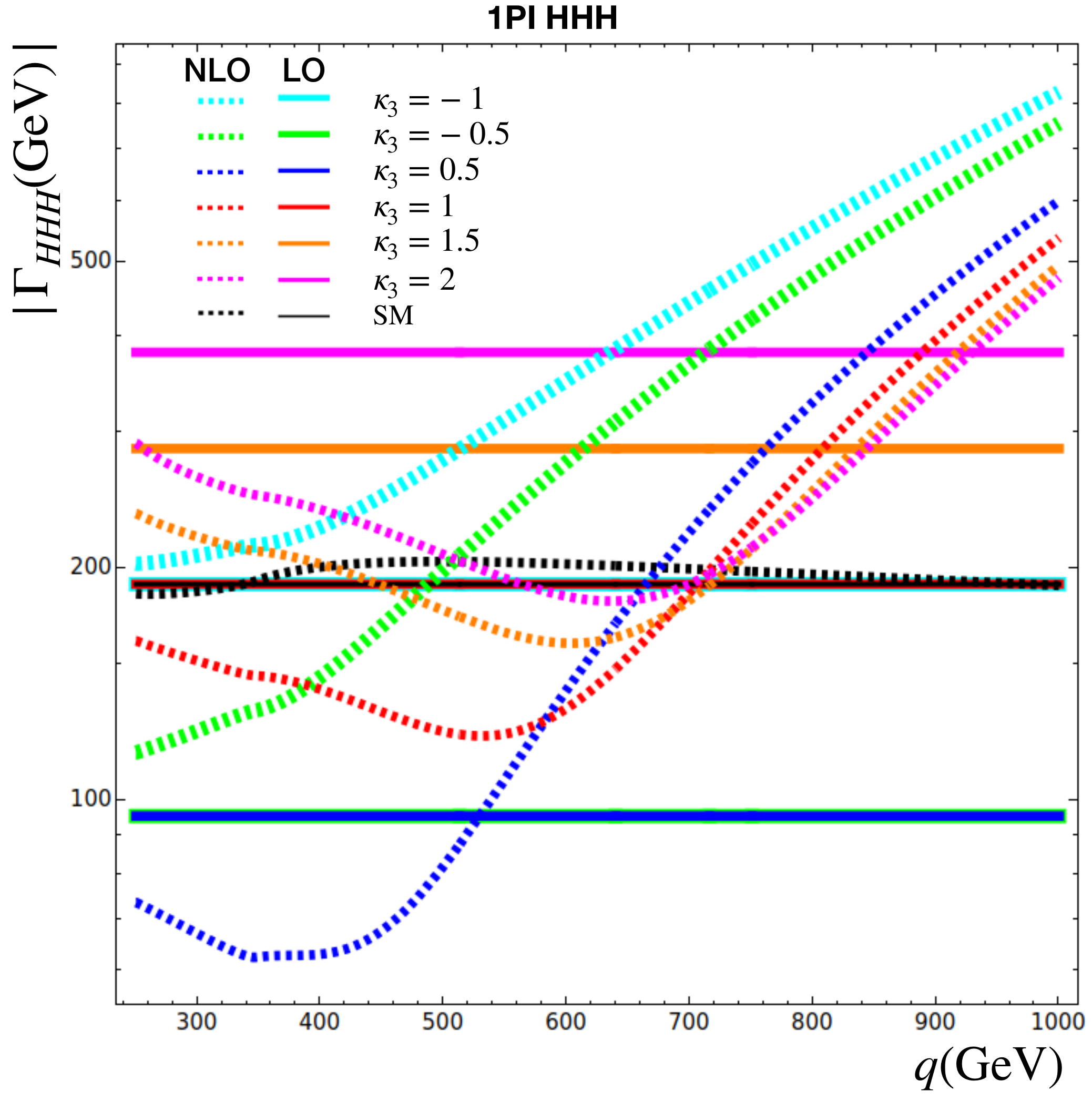}
\includegraphics[width=0.53\textwidth]{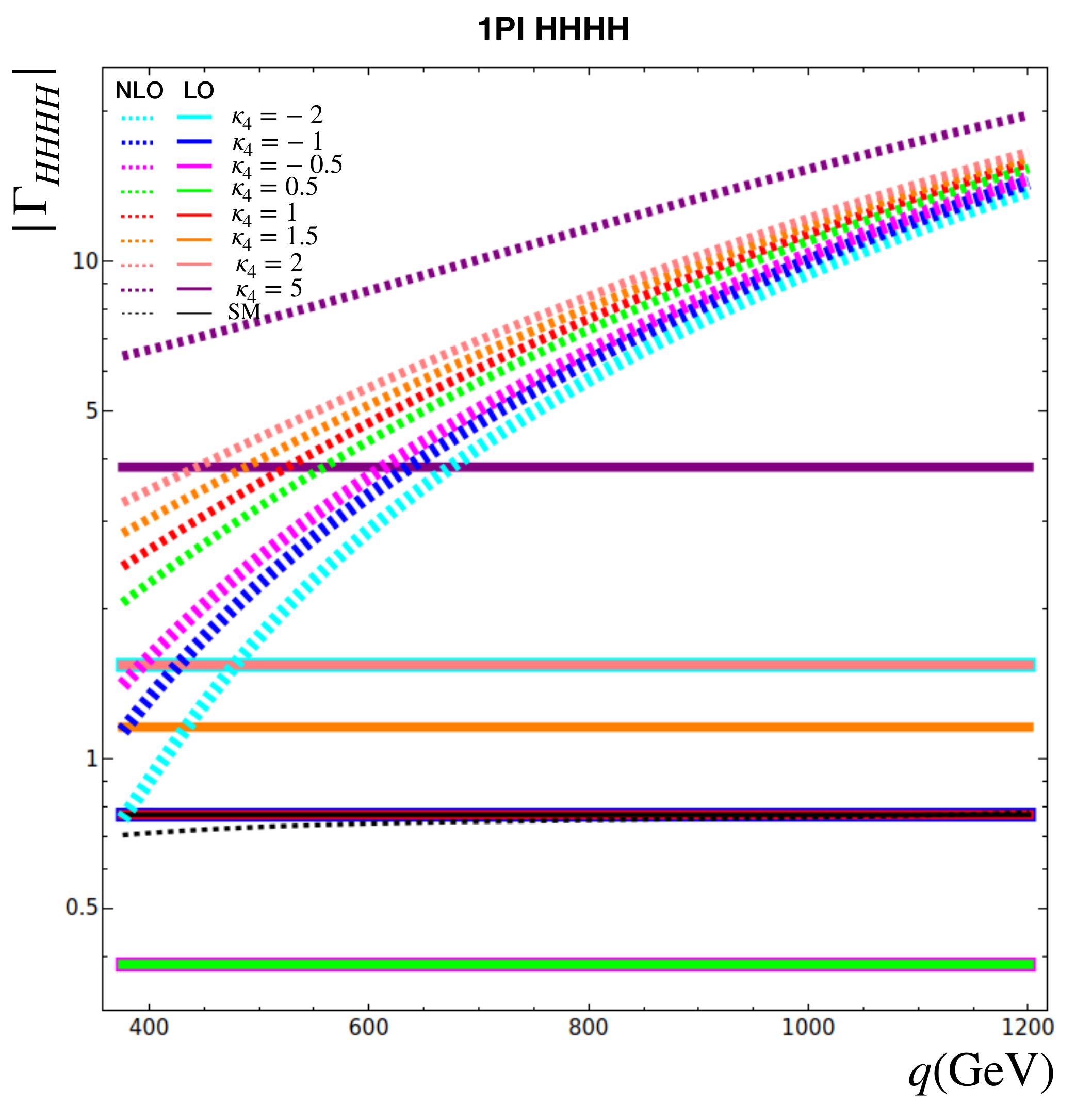}
\caption{Predictions for the 1PI HHH (left) and 1PI HHHH (right) functions within the HEFT, LO and NLO,  as a function of the off-shell momentum $q$ and for various values of the $\kappa_i$ parameters. ($\kappa_3$ in the left plot,  $\kappa_4$ in the left plot).  The other HEFT coefficients are set to $a=b=1$ and $a_i=0$.  The momenta assignement in HHH is $p_1,p_2,p_3$ with $p_1^2=q^2$ and $p_2^2=p_3^2=m_H^2$, and in HHHH is $p_1,p_2,p_3, p_4$ with $p_1^2=q^2$ and $p_2^2=p_3^2=p_4^2=m_H^2$. In the case of HHHH, we also fix the values $s_{23}=(p_2+p_3)^2=(1500 GeV)^2$ and $t=(p_1-p_3)^2=-(500 GeV)^2$ as an example. } 
\label{HHH-HHHH-q}
\end{figure}

\begin{figure}[!t]
\centering
\includegraphics[width=0.95\textwidth]{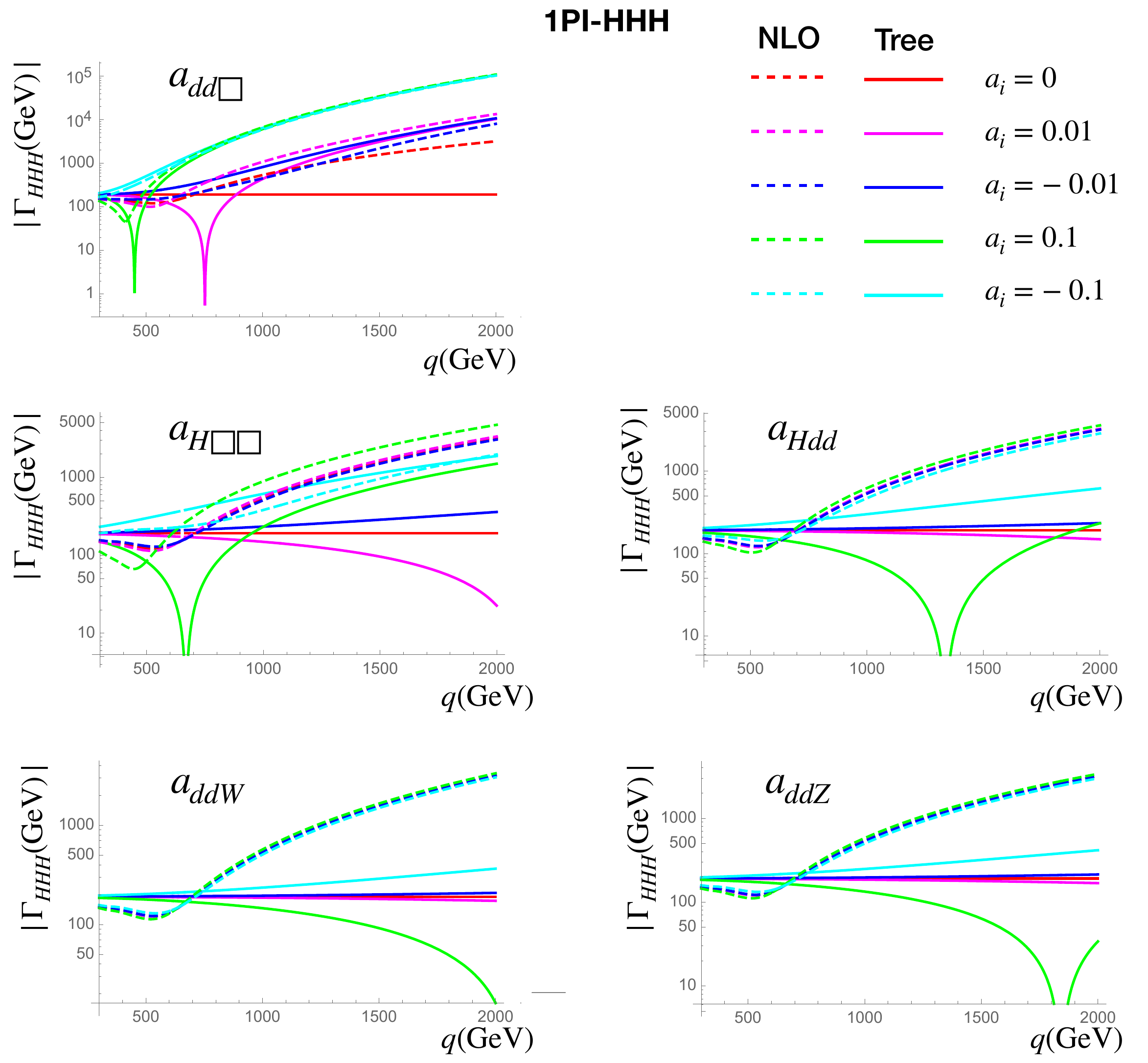}
\caption{Predictions for the 1PI HHH functions within the NLO HEFT as a function of the off-shell momentum $q$ and for various values of the relevant $a_i$ coefficients.  The LO HEFT coefficients are set to $a=b=\kappa_3=\kappa_4=1$.  The setting for the momenta are as in the previous figures.}
\label{HHH-ais}
\end{figure}

\begin{figure}[!t]
\centering
\includegraphics[width=0.95\textwidth]{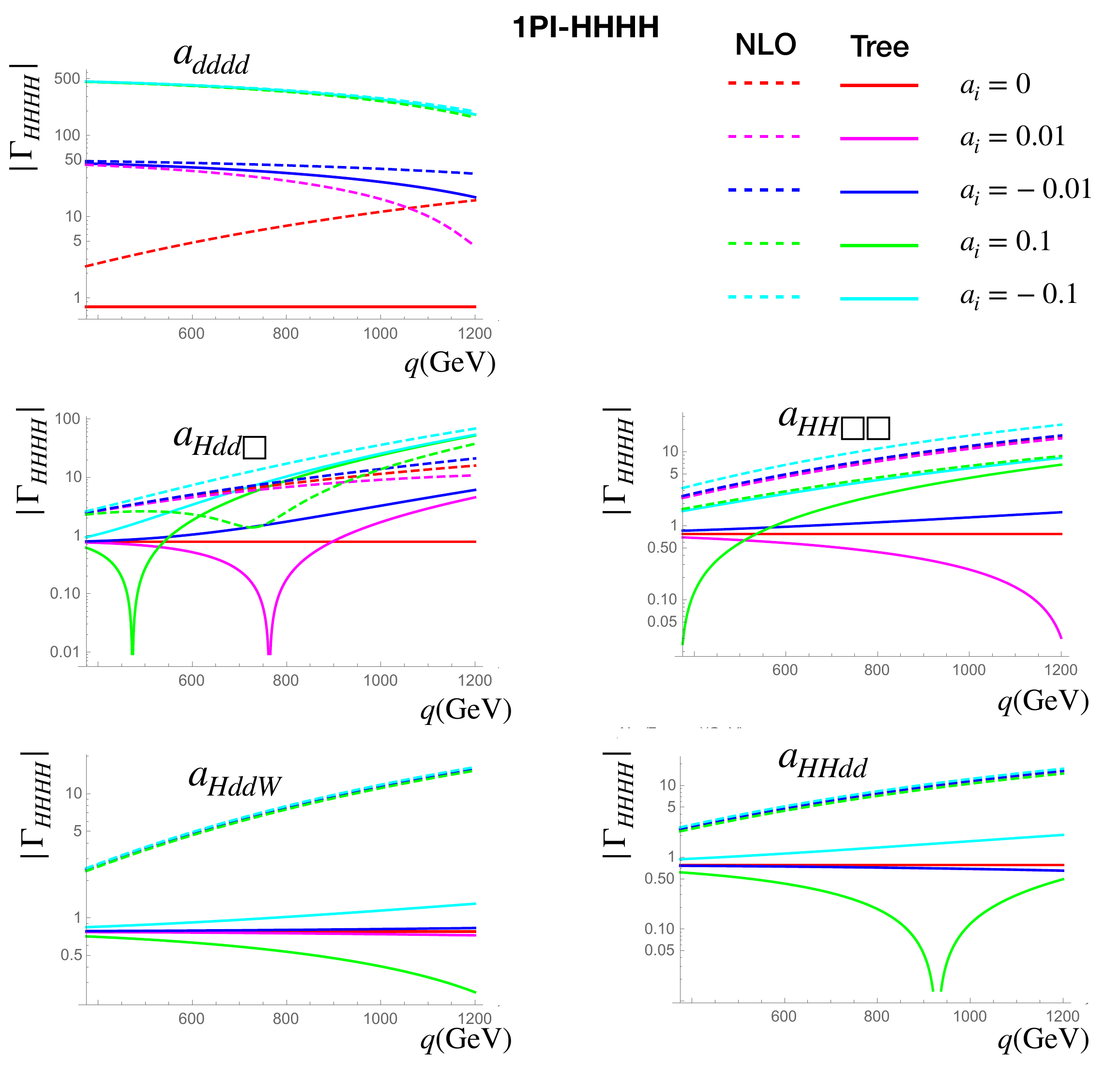}
\caption{Predictions for the 1PI HHHH functions within the NLO HEFT as a function of the off-shell momentum $q$ and for various values of the relevant $a_i$ coefficients.  The LO HEFT coefficients are set to $a=b=\kappa_3=\kappa_4=1$.  The setting for the momenta is as in the previous figures.  Notice that 
the plot for $ a_{HddZ}$  is not included here because it is practically identical to the $a_{HddW}$ one.}
\label{HHHH-ais}
\end{figure}

We summarize in Figs.~\ref{HH-q} through \ref{HHHH-ais} the most relevant numerical results for the $\greenfR^{\rm NLO}$ functions.  In this numerical analysis, we include both the radiative corrections from the bosonic sector and from the fermionic sector,  more specifically from top loops.   We have computed the top loop corrections to all the involved 1PI functions within the SM framework since we have assumed that the top couplings to gauge bosons $W$, $Z$, $\gamma$ (and $g$) and the Higgs boson $H$ are exactly as in the SM.  The top loops (not shown here for brevity) are just the sunset diagram in $HH$,  the triangle diagram in $HHH$ and the box diagram in $HHHH$.    On the other hand,  our main focus in this section is to show the main effects on the 1PI functions that produce relevant distant predictions between the HEFT and the SM.  These most relevant BSM effects are basically: 1) from the virtuality of the external legs,  2) from the LO HEFT coefficients $\kappa_3$ and $\kappa_4$ and 3) from the NLO HEFT $a_i$ coefficients. For simplicity,  we will show in this section the effect of just one external leg being off-shell whereas the others are set on-shell. Specifically,  in $H(p_1)H(p_2)H(p_3)$ we are assuming $p_1^2=q^2$,  $p_2^2=p_3^2=\mh^2$, and in $H(p_1)H(p_2)H(p_3)H(p_4)$ we are assuming $p_1^2=q^2$,  $p_2^2=p_3^2=p_4^2=\mh^2$.  In the $HH$ case, the unique momentum involved $q\equiv \sqrt{q^2}$ is obviously off-shell.  Finally,  as said in the text,  in all the numerical estimates in this work we take the input values of $a=b=1$. 

%%%%%%%%%%%%%%%%%%%%%%%%%%%%%%%%%%%%%%%%%%%%%%%%%%%%%
%%%%%%%%%%%%%%%%%%%%%%%%%%%%%%%%%%%%%%%%%%%%%%%%%%%%%

First, we show in Fig.~\ref{HH-q} the results for the modulus of the NLO renormalized self-energy as a function of the off-shell momentum $q$ defining the degree of virtuality in the internal Higgs boson that is propagating in the total process. The left plot is for $a_{\Box\Box}=0$ and we have included the predictions from the HEFT and from the SM for comparison. The contributions from just bosonic loops are included separately in order to compare them with the full results including both bosonic and top loops.   As we can see, the size of the top loop corrections in the HEFT case is much smaller than the bosonic ones and these latter dominate largely the $HH$ rates (the orange line is underlying the red line).  In contrast, in the SM case, the two contributions from bosonic and top loops are of similar size and therefore compete in the full result. Comparing the full HEFT and full SM results, it is clear that the effect from the virtualty of $q$ is more pronounced in the HEFT than in the SM and this can lead to large departures in the predicted HEFT rates compared to the SM rates.  When the effect of $a_{\Box\Box} \neq 0$ is incorporated in the full results (right plot), the departures of the HEFT prediction compared to the SM ones can be even more separated,  particularly for negative  $a_{\Box\Box}$.  As a consequence, the rates for the considered multi-Higgs processes at LHC can also be very different, as is seen in the text.

%%%%%%%%%%%%%%%%%%%%%%%%%%%%%%%%%%%%%%%%%%%%%%%%%%%%%
%%%%%%%%%%%%%%%%%%%%%%%%%%%%%%%%%%%%%%%%%%%%%%%%%%%%%

Fig.~\ref{HHH-HHHH-q} shows the HEFT predictions for the modulus of the 1PI (complex) functions $\Gamma_{HHH}$ (left) and $\Gamma_{HHHH}$ (right) as a function of the off-shell momentum $q$ and compare them with the SM predictions. We include the full predictions NLO (dashed lines) and LO(solid lines) in both cases HEFT (in colour) and SM (in black). The HEFT coefficients explored in these plots are  $\kappa_3$ and $\kappa_4$, for which we are setting some illustrative values shown in the legend.  The other HEFT coefficients $a_i$'s are set to zero. Notice that this setting of all $a_i=0$  implies $\Gamma^{\rm a_i} =0$ and therefore the tree level predictions and LO predictions coincide,  i.e.  $\Gamma^{\rm tree}= \Gamma^{\rm LO}$.  Notice also that in the case of the SM predictions, this identification is always true. 

%%%%%%%%%%%%%%%%%%%%%%%%%%%%%%%%%%%%%%%%%%%%%%%%%%%%%
%%%%%%%%%%%%%%%%%%%%%%%%%%%%%%%%%%%%%%%%%%%%%%%%%%%%%
  
Let us first comment on the $HHH$ case (left plot) The comparison of the LO HEFT predictions with the LO SM predictions (both constant with $q$)  shows the expected linear shift with $\kappa_3$.  For $\kappa_3=1$, both LO HEFT and LO SM predictions coincide.  For $|\kappa_3|>1$ ($|\kappa_3|<1$), the HEFT prediction is shifted upwards (downwards) compared to the SM.  We next compare the NLO predictions which now clearly show a different $q$ dependence. The NLO HEFT lines show a very important effect from the virtuality of $q$ separating the predictions from the corresponding LO ones.  This is in contrast to the NLO SM prediction which shows a very small departure compared to the LO SM one.  This demonstrates that the size of the radiative correction (highly dominated by the bosonic loops) within the HEFT can be large,  whereas within the SM it is very small.  It should also be noticed that this is true even at low $q$ close to the threshold value of $q=2m_H$.  For $\kappa_3=1$, we find different values for the NLO HEFT prediction and the NLO SM one even at this close to threshold region. This difference also reflects the non-linearity feature within the bosonic HEFT loops in comparison with the bosonic SM loops.  A summary of ratios NLO/LO will be given at the end of this appendix in Tab.~\ref{tab:ratios}. Finally, regarding the sign of the NLO correction compared to the LO prediction, we see that for the SM it is negative for low $q<350 \,  {\rm GeV}$ and positive for $q>350 \,  {\rm GeV}$.  In the HEFT case,  we find a positive correction for negative $\kappa_3$,  whereas for positive $\kappa_3$ we find a negative correction at low $q$ that changes to positive at high $q$.  This change of positive to negative correction happens at around $q=400  \,  {\rm GeV}$,   $q=550  \,  {\rm GeV}$,  $q=600  \,  {\rm GeV}$, $q=650  \,  {\rm GeV}$ for $\kappa_3=0.5,  1,  1.5, 2$, respectively.  The most relevant outcome from this plot is that the NLO HEFT prediction for this 1PI $HHH$  function defining the size of the $HHH$ blob vertex in the text could lead to large departures compared to the SM values,  showing both possibilities either increasing or decreasing the SM rates.  We have also explored the $\kappa_4$ effect via the loops in the $HHH$ case but we do not show it here because it is less relevant being very small.
 
Regarding the $HHHH$ case (right plot), the most relevant parameter is $\kappa_4$. The $\kappa_3$ parameter also enters via the loops but its effects are very tiny and less relevant for the present computation.  Consequently,  they are not shown here.   As can be seen in this plot,  we have also found important departures between the HEFT and the SM predictions. The comparison of the LO HEFT and LO SM predictions shows the expected linear shift with $|\kappa_4|$, being upwards for  $|\kappa_4|>1$ and downwards for $|\kappa_4|<1$.  Going NLO notice again that the HEFT prediction for $\kappa_4=1$ (dashed red line)  does not coincide with the SM prediction (dashed black line) and the distance is large even at low $q$ close to the threshold value of $q=3m_H$.  The size of the NLO correction in the SM case is again very small compared to the size of the correction within the HEFT.  As for the sign of these corrections,  we find them negative in the SM case and positive in all the HEFT cases studied, except for the largest negative value considered in this plot of $\kappa_4=-2$ where the correction is negative for $q<470 \,{\rm GeV}$ and positive for  $q >470 \,{\rm GeV}$. 
 In any case,   the most relevant effect in this 1PI function as in the previous case is the off-shell effect that can change the size of the $HHHH$ blob vertex considerably.

%%%%%%%%%%%%%%%%%%%%%%%%%%%%%%%%%%%%%%%%%%%%%%%%%%%%%
%%%%%%%%%%%%%%%%%%%%%%%%%%%%%%%%%%%%%%%%%%%%%%%%%%%%%

Finally, we show next in Fig.~\ref{HHH-ais} and in Fig.~\ref{HHHH-ais} the effects from the NLO HEFT $a_i$ coefficients on the 1PI $HHH$ and $HHHH$ functions, respectively. In these plots, we set  $\kappa_3=\kappa_4=1$. In this case, we show again the results as a function of the virtuality of $q$ and choose several benchmark values for each single non-vanishing $a_i$ coefficient (the other $a_i$'s are set to zero).  Concretely,  we explore the values $a_i=0, \pm 0.01, \pm 0.1$.  We include both the tree predictions,  $\Gamma^{\rm tree}$ (solid lines), and the full one-loop ones,  $\greenfR^{\rm NLO}$ (dashed lines).  Our main interest here is to learn on the relative size of the radiative corrections, i.e. $\mO(\hbar^1)$  versus $\mO(\hbar^0)$ and the relative importance of the various $a_i$'s coefficients.   Notice that,  in this case,  the tree HEFT and the tree SM predictions coincide for the particular setting  $a_i=0$ (solid red lines).  
 
%%%%%%%%%%%%%%%%%%%%%%%%%%%%%%%%%%%%%%%%%%%%%%%%%%%%%
\begin{table}
\centering
\begin{tabular}{|c|c|c||c|c|c|}
\cline{1-3}\cline{4-6}
\multicolumn{3}{|c||}{$|\Gamma_{HHH}^{NLO}|/|\Gamma_{HHH}^{LO}|$} & \multicolumn{3}{|c|}{$|\Gamma_{HHHH}^{NLO}|/|\Gamma_{HHHH}^{LO}|$} \rule[-.3cm]{0pt}{5ex}\\\hline
$\kappa_3$ & $q=251$ GeV & $q=1000$ GeV & $\kappa_4$ & $q=376$ GeV & $q=1000$ GeV \\
\hline
-1 & 1.1 & 4.4 & -2 & 0.49 & 6.2 \\
-0.5 & 1.2 & 7.9 & -1 & 1.5 & 13 \\
0.5 & 0.77 & 6.3 & -0.5 & 3.7 & 27 \\
1 & 0.84 & 2.8 & 0.5 & 5.4 & 29 \\
1.5 & 0.82 & 1.7 & 1 & 3.2 & 15 \\
2 & 0.76 & 1.3 & 1.5 & 2.5 & 10 \\
& & & 2 & 2.1 & 7.9 \\
& & & 5 & 1.7 & 4.0 \\\hline
SM & 0.97 & 1.0 & SM & 0.91 & 0.99 \\
\hline
\end{tabular}
\caption{NLO/LO ratios for the  1PI HHH and 1PI  HHHH functions for different values of $\kappa_3$ and $\kappa_4$,  respectively.  Two values of $q$ are chosen in each case.   The setting for the other external momenta involved in the 1PI functions are defined  as in the previous figures. The SM predictions are  also included for comparison.}
\label{tab:ratios}
\end{table}

%%%%%%%%%%%%%%%%%%%%%%%%%%%%%%%%%%%%%%%%%%%%%%%%%%%%%

Let us first comment on the 1PI $HHH$ results in Fig.~\ref{HHH-ais}.  We notice the appearance of dips in some solid lines at specific $q$ values,  which are produced because of the two competing contributions to the tree predictions,  $\Gamma^{\rm LO}$ which is $q$ independent and  $\Gamma^{\rm a_i}$ which is $q$ dependent.  Depending on the sign of the  $a_i$'s these two contributions may interfere constructively/destructively (add or subtract) and produce cancellations. On the other hand,  in the large $q$ region, it is clear that the most relevant $a_i$ coefficient is  $a_{dd \Box}$ since it leads to larger values of $\Gamma_{HHH}$ as compared to other coefficients with the same size. For instance, for  $a_{dd \Box} = 0.1$ we get at $q=1000 \, {\rm GeV}$ a value of  $\Gamma_{HHH} \sim 6000 \, {\rm GeV}$,  whereas for the other coefficients with the same size we get a smaller prediction $\Gamma_{HHH} < 1000 \, {\rm GeV}$.  Another relevant coefficient at large $q$ is  $a_{H\Box \Box}$.  In general, we find important radiative corrections in several cases (depending on the $q$ and $a_i$ values) manifesting as large separations between the NLO and the tree predictions. In summary, the largest HEFT departures compared to the SM case are obtained at large virtuality of $q$ and large $|a_i|$ values. Similar conclusions are obtained from the 1PI $HHHH$ results in Fig.~\ref{HHHH-ais}.  In this case,  the most relevant parameters at large $q$ values seem to be $a_{dddd}$ and $a_{Hdd\Box}$,  since they provide the largest deviations compared to the SM rates. 

In summary,  when going from LO to NLO within the HEFT,  we find important corrections in the 1PI $HHH$ and $HHHH$ functions defining the size of the blob vertices involved in the $HH$ and $HHH$ production from gluon-gluon fusion at the LHC.  These corrections depend notably on the size of the virtuality of the internal propagating Higgs boson momentum and can be large depending on the $q$ values.  Obviously,  for the relevant estimates at the LHC, the $q$ values are finally integrated over all the available phase space.  If we focus on the sensitivity to $\kappa_3$ and $\kappa_4$,  we summarize in Tab.~\ref{tab:ratios} the predicted ratios for $|\Gamma^{\rm NLO}_{HHH}/\Gamma^{\rm LO}_{HHH}|$ and  $|\Gamma^{\rm NLO}_{HHHH}/\Gamma^{\rm LO}_{HHHH}|$ at two different values of $q$,  and for several benchmark values of $\kappa_3$ and $\kappa_4$.  The two selected values for the virtuality  are 1) $q$ close to the threshold production value,  i.e. $q\simeq 2m_H$  in $HH$ production and  $q \simeq 3m_H$ in $HHH$ production,  and 2) $q=1000\, {\rm  GeV}$. For instance,  for the lowest $q$ values close to threshold values we find that the size of the radiative corrections in the 1PIs within the HEFT  (depending on the $\kappa_i$ values) are: 1) $10\%-24\%$ for $HHH$ and 2) $ \gtrsim 50\%$ for $HHHH$.  These radiative corrections should be compared with the corresponding SM radiative corrections which are small as deduced from this table,  $ \sim 3\%$ for $HHH$ and $\sim 9\%$ for $HHHH$. They are in concordance with the final cross section NLO/LO rates presented in the text, in particular in Fig.~{\ref{fig:kfac}}. Obviously, to understand the comparison with the relevant $HH$ and $HHH$ production rates at the LHC,  these ratios of 1PI vertices that depend on $q$ should be integrated over all available LHC phase space. But they provide in any case a useful reference.  

%%%%%%%%%%%%%%%%%%%%%%%%%%%%%%%%%%%%%%%%%%%%%%%%%%%%%
%  bibliography
%%%%%%%%%%%%%%%%%%%%%%%%%%%%%%%%%%%%%%%%%%%%%%%%%%%%%
%\bibliographystyle{JHEP}
\bibliography{paper}
%%%%%%%%%%%%%%%%%%%%%%%%%%%%%%%%%%%%%%%%%%%%%%%%%%%%%  

\end{document}